\documentclass[sigconf,authorversion,nonacm]{acmart}
\AtBeginDocument{%
  }

\usepackage[nolist]{acronym}  
\usepackage[T1]{fontenc}
\usepackage{calc}
\usepackage{multirow}
\usepackage{tikz}
\usetikzlibrary[shapes.geometric]
\usetikzlibrary{shapes.misc}
\usetikzlibrary{calc}
\usepackage{makecell}

\usepackage{booktabs} 
\usepackage{harveyballs}
\usepackage{collcell}

\usepackage{xcolor}

\usepackage{colortbl}
\usepackage{array}
\usepackage{tabularx}
\usepackage{adjustbox}
\usepackage{diagbox}

\usepackage[most]{tcolorbox}

\usepackage{adjustbox}
\usepackage{caption}
\usepackage{subcaption}
\usepackage{pgfplots}
\usepackage{pgfplotstable}

\usepackage{soul}
\usepackage{tabularx}

\pgfplotsset{compat=1.18} 
\usepackage{circledtext}
\usepackage{pifont}
\usepackage{fontawesome5} 

\usepackage{url}
\usepackage{hyperref}
\hypersetup{hidelinks}

\usepgfplotslibrary{groupplots}
\pgfplotsset{compat=newest}

\usepackage{stackengine}

\newcolumntype{L}[1]{>{\raggedright\let\newline\\\arraybackslash\hspace{0pt}}m{#1}}
\newcolumntype{C}[1]{>{\centering\let\newline\\\arraybackslash\hspace{0pt}}m{#1}}
\newcolumntype{R}[1]{>{\raggedleft\let\newline\\\arraybackslash\hspace{0pt}}m{#1}}
\usepackage[inline]{enumitem}

\definecolor{risk1}{rgb}{0.1, 0.8, 0.1} 

\definecolor{risk2}{rgb}{0.5, 1.0, 0.5}  
\definecolor{risk3}{rgb}{1.0, 1.0, 0.5}  
\definecolor{risk4}{rgb}{1.0, 0.5, 0.0}      
\definecolor{risk5}{rgb}{1.0, 0.0, 0.0}     

\newcommand{\ApplyGradient}[1]{%
    \ifnum #1=1 
        \colorbox{risk1}{\textbf{#1}}%
    \fi
     \ifnum #1=2 
        \colorbox{risk2}{\textbf{#1}}%
    \fi
     \ifnum #1=3 
        \colorbox{risk3}{\textbf{#1}}%
    \fi
     \ifnum #1=4
        \colorbox{risk4}{\textbf{#1}}%
    \fi
     \ifnum #1=5 
        \colorbox{risk5}{\textbf{#1}}%
    \fi
}

\newlength{\var}
\newcommand*{\BlackCircle}[0]{\tikz\draw[black,fill=black] (0,0) rectangle (1.5ex,\var);}
\newcommand*{\WhiteCircle}[0]{\tikz\draw[black,fill=white] (0,0) rectangle (1.5ex,1\var);}
\newcommand*{\GrayCircle}[0]{\tikz\draw[black,fill=gray!40] (0,0) rectangle (1.5ex,\var);}

\newcommand{\VeryLowFeasibility}[0]{%
    \tikz[baseline=-0.6ex]{
        \draw[ fill=white] (0,0) circle (0.8ex);
        \fill[green] (0,0) -- (0.8ex,0) arc[start angle=0,end angle=90,radius=0.8ex] -- cycle;
        \draw[] (0,0) -- (0.8ex,0); 
        \draw[] (0,0) -- (0,0.8ex); 
    }%
}

\newcommand{\LowFeasibility}[0]{%
    \tikz[baseline=-0.6ex]{
        \draw[fill=white] (0,0) circle (0.8ex);
        \fill[green!10!yellow] (0,0) -- (0,-0.8ex) arc[start angle=-90,end angle=90,radius=0.8ex] -- (0,0) -- cycle;
        \draw[] (0,0) -- (0,0.8ex); 
        \draw[] (0,0) -- (0,-0.8ex); 
    }%
}

\newcommand{\MediumFeasibility}[0]{%
    \tikz[baseline=-0.6ex]{
        \draw[fill= white] (0,0) circle (0.8ex);
        \fill[fill=orange!70] (0,0) -- (-0.8ex,0) arc[start angle=-180,end angle=90,radius=0.8ex] -- cycle;
        \draw[] (0,0) -- (-0.8ex,0); 
        \draw[] (0,0) -- (0,0.8ex); 
    }%
}

\newcommand{\HighFeasibility}[0]{%
    \tikz[baseline=-0.6ex]{
        \draw[fill= white] (0,0) circle (0.8ex);
        \fill[fill=red] (0,0) -- (-0.8ex,0) arc[start angle=-180,end angle=180,radius=0.8ex] -- cycle;
    }%
}

\newcommand{\targeted}{
    \resizebox{0.8em}{!}{%
    \tikz[baseline=-10ex]{
        \fill[black] (0,0) circle (1.4cm);
		\fill[white] (0,0) circle (1.1cm);
		\fill[black] (0,0) circle (0.8cm);
		\fill[green] (0,0) circle (0.5cm);

		\draw[draw=black, fill=black] (-0.1cm, 1.3cm) rectangle ++(0.1cm, 0.3cm);
		\draw[draw=black, fill=black] (1.3cm, -0.1cm) rectangle ++(0.3cm, 0.1cm);
		\draw[draw=black, fill=black] (-0.1cm, -1.3cm) rectangle ++(0.1cm, -0.3cm);
		\draw[draw=black, fill=black] (-1.3cm, -0.1cm) rectangle ++(-0.3cm, 0.1cm);
    }
    }
}

\newcommand{\ntargeted}{
    \resizebox{0.8em}{!}{%
    \tikz[baseline=-10ex]{
        \fill[black] (0,0) circle (1.4cm);
		\fill[white] (0,0) circle (1.1cm);
		\fill[black] (0,0) circle (0.8cm);
		\fill[green] (0,0) circle (0.5cm);

		\draw[draw=black, fill=black] (-0.1cm, 1.3cm) rectangle ++(0.1cm, 0.3cm);
		\draw[draw=black, fill=black] (1.3cm, -0.1cm) rectangle ++(0.3cm, 0.1cm);
		\draw[draw=black, fill=black] (-0.1cm, -1.3cm) rectangle ++(0.1cm, -0.3cm);
		\draw[draw=black, fill=black] (-1.3cm, -0.1cm) rectangle ++(-0.3cm, 0.1cm);
		
		\draw[draw=red, fill=red, rotate =-45] (-1.8cm,-0.1cm) rectangle ++(3.6cm, 0.2cm);
    }
    }
}

\newcommand{\prof}{
    \tikz[baseline=-0.5ex]{
        \node[anchor=center, inner sep=0pt] at (0,0) {\includegraphics[width=\var]{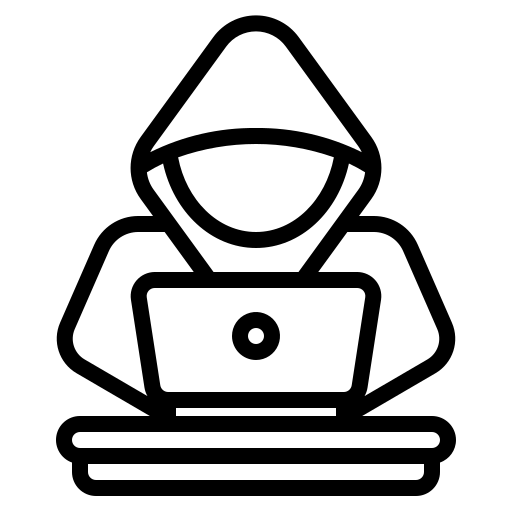}};
    }
}
\newcommand{\expert}{
    \tikz[baseline=-0.5ex]{
        \node[anchor=center, inner sep=0pt] at (0,0) {\includegraphics[width=\var]{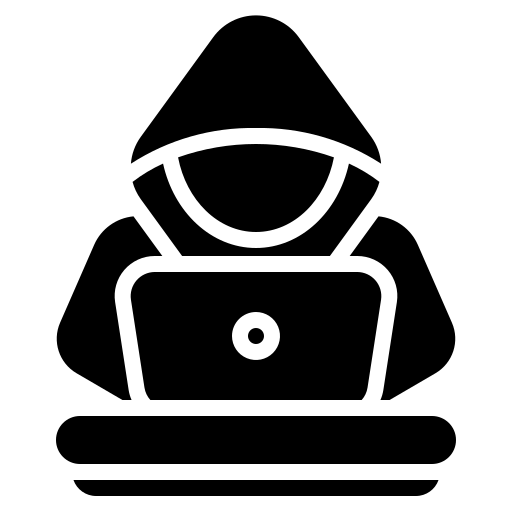}};
    }
}
\newcommand{\Layman}{
    \tikz[baseline=-0.5ex]{
        \node[anchor=center, inner sep=0pt] at (0,0) {\includegraphics[width=\var]{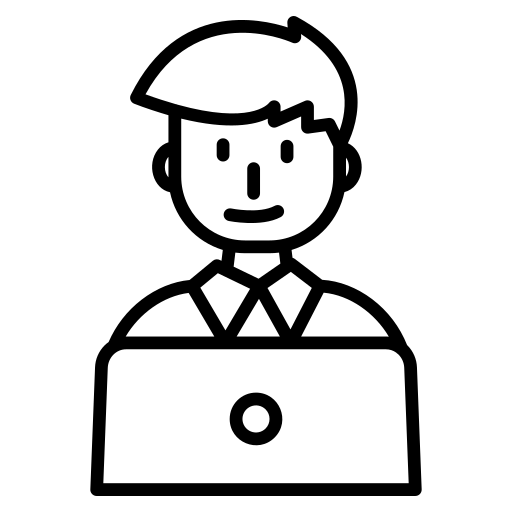}};
    }
}
\newcommand{\mexpert}{
    \tikz[baseline=-0.5ex]{
        \node[anchor=center, inner sep=0pt] at (0,0) {\includegraphics[width=\var]{img/symbol/expert.png}};
         \node[anchor=center, inner sep=0pt] at (\var,0) {\includegraphics[width=\var]{img/symbol/expert.png}};
    }
}

\definecolor{negligible}{rgb}{0.0, 0.5, 0.0} 
\definecolor{moderate}{rgb}{0.5, 1.0, 0.5}  
\definecolor{major}{rgb}{1.0, 0.5, 0.0}      
\definecolor{severe}{rgb}{1.0, 0.0, 0.0}     

\newcommand{\impact}[1]{
    \begin{tikzpicture}
        \foreach \i in {1,2,3,4} {
            \ifnum\i=#1
                \ifnum\i=1 \fill[negligible] (0,0) rectangle (1ex, \var); \fi
                \ifnum\i=2 \fill[moderate] (1ex,0) rectangle (2ex, \var); \fi
                \ifnum\i=3 \fill[major] (2ex,0) rectangle (3ex, \var); \fi
                \ifnum\i=4 \fill[severe] (3ex,0) rectangle (4ex, \var); \fi
            \fi
            \draw (1ex*\i,0) -- (1ex*\i,\var); 
              \draw (0,0) rectangle (4ex, \var);
        }
    \end{tikzpicture}
}

\newcommand{\circledtexted}[1]{%
  \raisebox{.5pt}{\textcircled{\raisebox{-.9pt}{\hspace{2pt}\smash{#1}\hspace{2pt}}}}%
}

\newcommand{\triangledtextxx}[2][0pt]{%
  \circledtextset{height=0.3cm}
  \raisebox{#1}{\circledtext[charf=\rmfamily\huge, resize=real, boxtype=O, boxfill=white, charshrink=0.8]{\uppercase\expandafter{\romannumeral#2\relax}}}
}

\newcommand{\triangledtextdd}[2][0pt]{%
  \circledtextset{height=0.3cm}
    \setlength{\fboxsep}{0pt}
    \raisebox{#1}{\hspace{1pt}\fbox{\parbox[c][1em][c]{1em}{\centering\scriptsize\uppercase\expandafter{\romannumeral #2\relax}}}\hspace{1pt}}%
}

\newcommand{\triangledtext}[2][0pt]{%
  \circledtextset{height=0.3cm}
    \setlength{\fboxsep}{0pt}
    \raisebox{#1}{\hspace{1pt}\raisebox{0.2ex}{\fbox{\parbox[c][1em][c]{1em}{\centering\scriptsize\uppercase\expandafter{\romannumeral #2\relax}}}}\hspace{1pt}}%
}

\newcommand{\triangledtextxxx}[2][0pt]{%
    \circledtextset{height=0.3cm}
    \setlength{\fboxsep}{0pt}
    \fbox{\parbox[c][1em][c]{1em}{\centering\scriptsize\uppercase\expandafter{\romannumeral #2\relax}}}%
}

\NewDocumentCommand{\progressbar}{O{1}m}{%
    \rotatebox[origin=c]{90}{%
    \begin{tikzpicture}[baseline=-1pt]
        
        \pgfmathsetmacro{\progress}{#2}
        \ifdim \progress pt < 0.26pt
            \definecolor{progresscolor}{rgb}{0,0.5,0} 
        \else
            \ifdim \progress pt < 0.51pt
                \definecolor{progresscolor}{rgb}{0,1,0} 
            \else
                \ifdim \progress pt < 0.76pt
                    \definecolor{progresscolor}{rgb}{1,0.65,0} 
                \else
                    \definecolor{progresscolor}{rgb}{1,0,0} 
                \fi
            \fi
        \fi
        \draw[fill= progresscolor, progresscolor ] (0,0) rectangle (#1 * \progress, 0.2);
        \draw[thin, anchor=south] (0,0) rectangle (#1, 0.2);
    \end{tikzpicture}
    }
}

\newcommand{\OK}{
    \tikz[baseline=-0.5ex]{
        \node[anchor=center] at (0,0) {\includegraphics[width=2.8ex]{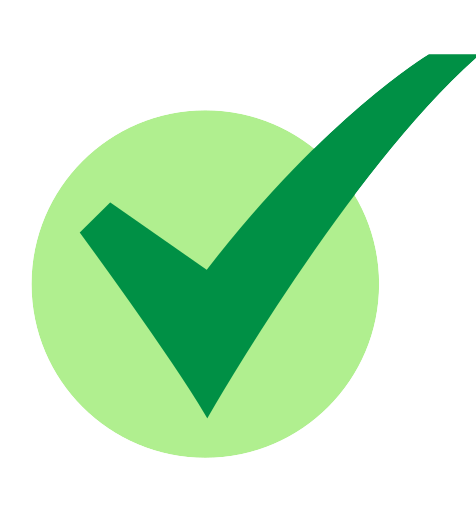}};
    }
}

\newcounter{observationcounter}
\newtcolorbox{observationbox}{
    enhanced,
    boxrule=0pt,                   
    borderline west={4pt}{0pt}{green!75!black}, 
     colframe=white,         
    boxrule=0pt,            
    colback=green!5!white,        
    sharp corners,                 
    left=4pt,                      
    right=4pt,                     
}

\NewDocumentCommand{\observation}{m}{%
    \vspace{-.5em}
    \stepcounter{observationcounter}
    \begin{observationbox}
        \textbf{Observation \theobservationcounter:} #1
    \end{observationbox}
    \vspace{-.5em}
}

\newcounter{researchgapcounter}
\newtcolorbox{Gapgbox}{
    enhanced,
    boxrule=0pt,                   
    borderline west={4pt}{0pt}{red!55!white}, 
    colframe=white,         
    boxrule=0pt,            
    colback=red!05!white,        
    sharp corners,                 
    left=4pt,                      
    right=4pt,                     
}

\newcommand{\afterfigspace}{\vspace{0cm}}

\newcommand{\cpar}[1]{{\vspace*{0.2\baselineskip}\noindent\bfseries #1:\ }}

\newcommand{\secref}[1]{\S\ref{#1}}

\newcommand{\cmark}{\ding{51}}

\newcommand{\analysis}{\faSearch}       
\newcommand{\optimization}{\faWrench}   
\newcommand{\updatei}{\faSync}           

\newcommand{\Cline}[1]{  
    \arrayrulecolor{gray!30}\cline{#1}\arrayrulecolor{black}
}

\acmYear{2026}
\acmConference[ASIA CCS '26]{ACM Asia Conference on Computer and Communications Security}{June 01--05, 2026}{Bangalore, India}
\acmBooktitle{ACM Asia Conference on Computer and Communications Security (ASIA CCS '26), June 01--05, 2026, Bangalore, India}
\acmPrice{}
\acmDOI{10.1145/3779208.3785267}
\acmISBN{979-8-4007-2356-8/2026/06}

\begin{document}

\title[SoK: Security of the Image Processing Pipeline for Camera-based Sensing in Autonomous Vehicles]{SoK: Security of the Image Processing Pipeline\\ for Camera-based Sensing in Autonomous Vehicles}


\author{Michael K\"uhr}
\email{michael.kuehr@tum.de}
\affiliation{%
  \institution{Technical University of Munich}
  \city{Munich}
  \state{}
  \country{Germany}
}

\author{Mohammad Hamad}
\email{mohammad.hamad@tum.de}
\affiliation{%
  \institution{Technical University of Munich}
  \city{Munich}
  \state{}
  \country{Germany}
}

\author{Pedram MohajerAnsari}
\email{pmohaje@clemson.edu}
\affiliation{%
	\institution{Clemson University}
	\city{Clemson}
	\state{South Carolina}
	\country{USA}
}

\author{Mert D. Pes\'e}
\email{mpese@clemson.edu}
\affiliation{%
  \institution{Clemson University}
  \city{Clemson}
  \state{South Carolina}
  \country{USA}
}

\author{Sebastian Steinhorst}
\email{sebastian.steinhorst@tum.de}
\affiliation{%
  \institution{Technical University of Munich}
  \city{Munich}
  \state{}
  \country{Germany}
}

\renewcommand{\shortauthors}{K\"uhr et al.}

\begin{abstract}
Cameras are crucial sensors for autonomous vehicles. They capture images that are essential for many safety-critical tasks. To process these images, a complex pipeline with multiple layers is used. Security attacks on this pipeline can severely affect passenger safety and system performance. However, many attacks presented in scientific literature overlook the fact that there are different layers and, hence, the feasibility and impact of these attacks can vary. While there has been research to improve the quality and robustness of the image processing pipeline, these efforts are often orthogonal to security research without exploiting potential overlap and synergies. In this work, we aim to bridge this gap by combining security and robustness research for the image processing pipeline in autonomous vehicles. We thoroughly investigated the body of literature on the security and robustness of the image processing pipeline and selected 92 papers for deeper discussion in this SoK. For the security domain, we classify the risk of attacks using the automotive security standard ISO 21434, emphasizing the need to consider all layers for overall system security. With our online tool TARA-CAM, we propose an interactive method to perform threat analysis and risk assessment following the ISO standard. We also demonstrate how existing robustness research can help mitigate the impact of attacks, addressing the current research gap. Finally, we present PICT, an embedded open-source testbed that can influence various parameters across all layers, allowing researchers to analyze the effects of different defense strategies and attack impacts. With this SoK, we contribute a comprehensive discussion and systematic analysis of existing approaches to image processing pipeline security and robustness, together with an open-source tool and testbed that jointly facilitates hardening the image processing pipeline against existing and future security attacks.
\end{abstract}




\maketitle

\begin{acronym}
\acro{ECU}{Electronic Control Unit}
\acro{SoK}{Systematization of Knowledge}
\acro{CV}{Computer Vision}
\acro{ML}{Machine Learning}
\acro{CMOS}{Complementary metal–oxide–semiconductor}
\acro{CCD}{Charge-coupled device}
\acro{CFA}{color filter array}
\acro{HDR}{High Dynamic Range}
\acro{CSI-2}{Camera Serial Interface 2}
\acro{ISP}{Image Signal Processor}
\acro{LED}{Light-Emitting Diode}
\acro{DNG}{Digital Negative}
\acro{GUI}{Graphical User Interface}
\acro{TARA}{Threat Analysis and Risk Assessment}
\end{acronym}
\section{Introduction}

\begin{figure}[t!]
\begin{center}
  \includegraphics[width=\linewidth]{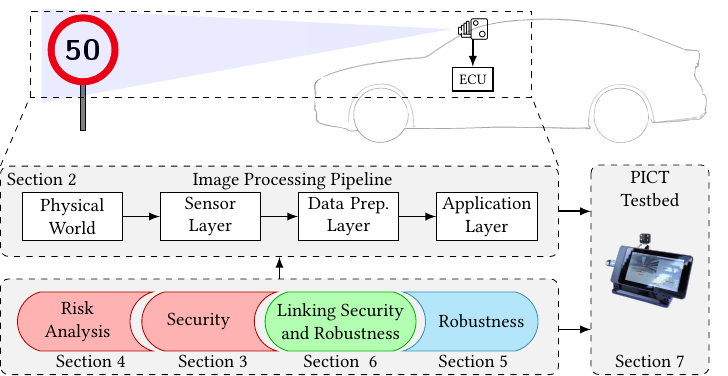}
  \caption{Abstraction of the image processing pipeline of autonomous vehicles and the main contribution of our \acs{SoK}.}
  \afterfigspace
  \label{fig:contribution}
\end{center}
\end{figure}

Autonomous vehicles and emerging advanced driver assistance systems have become a reality in recent years~\cite{mercedes-benz_group_ag_mercedes-benz_2023, waymo_llc_self-driving_2024}. These vehicles depend on sensing systems to interpret their surroundings and make real-time decisions. Many autonomous vehicles integrate LiDAR or radar, but camera systems are widely used as single information source for tasks such as traffic sign detection, traffic light recognition, and lane detection~\cite{ibanez-guzman_lidar_2025, gao_autonomous_2022} due to their ability to capture color information of the environment. Attacks on other sensors and sensor fusion have been examined in prior research~\cite{sato_lidar_2024, solanki_survey_2025, cao_invisible_2021}, but these do not replace the need for a detailed study of camera pipelines. Because cameras are often a primary or sole perception sensor, securing their image processing pipeline is critical, both for camera-only platforms and for multi-sensor systems where compromised visual input can still mislead fused perception. The high information density and environmental exposure of cameras enlarge the attack surface~\cite{yan_can_2016, el-rewini_cybersecurity_2020, gao_autonomous_2022, pham_survey_2021, liu_seeing_2021, boloor_attacking_2020}, posing severe risks, especially in fully autonomous vehicles without human fallback. Camera-based systems face diverse attacks that exploit both hardware and algorithmic weaknesses, including adversarial physical perturbations that mislead object detection~\cite{akhtar_advances_2021, guesmi_physical_2023, wei_visually_2023, wei_physical_2024, wang_survey_2023, wang_does_2023}, blinding attacks using directed light~\cite{petit_remote_2015, wang_i_2021}, and exploits targeting sensor-specific characteristics~\cite{sayles_invisible_2021, kohler_they_2021, li_light_2020}. These threats have been shown to be particularly effective against \ac{ML} algorithms~\cite{szegedy_intriguing_2014} that are often used in autonomous vehicles, and their effectiveness often depends on how they interact with different stages of the image processing pipeline. Analyzing and systematizing security and robustness research across this pipeline is the focus of our work.

\begin{figure*}[th!]
\begin{center}
  \includegraphics[width=0.9\linewidth]{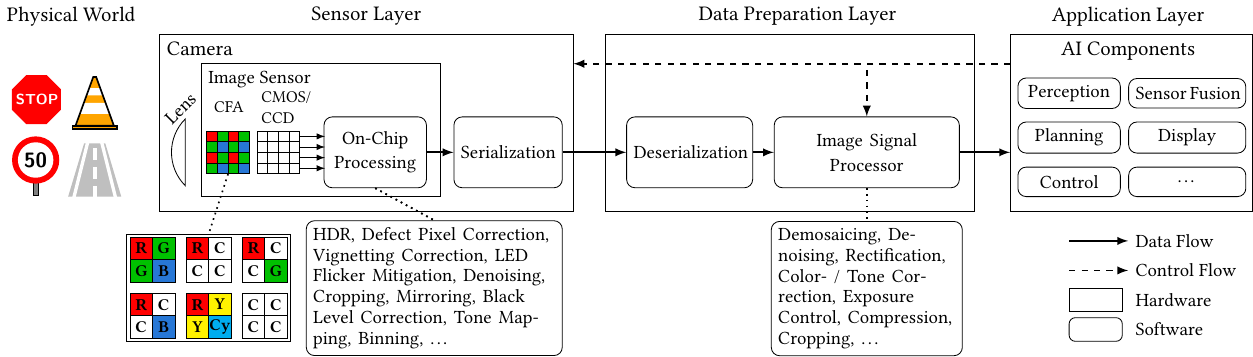}
  \caption{Overview of the image processing pipeline, consisting of four layers. Hardware components like the image sensor have solid edges, while software-configurable components have rounded corners, such as the \acl{ISP}.}
  \afterfigspace
  \label{fig:teaser}
\end{center}
\end{figure*}

Image capturing is not a simple, monolithic process but rather involves a complex, multi-stage pipeline. \autoref{fig:teaser} illustrates this complexity, showing how data flows from the physical world to \ac{CV} algorithms in the application layer (details on the pipeline are given in \secref{sec:detailsImagePipeline}). Ensuring the security of the image processing pipeline requires a deep understanding of its operational principles and a thorough identification of potential attack vectors. Our \ac{SoK} reveals a significant gap in the current research landscape: While well-known existing work focuses narrowly on the application layer, such as by modifying \ac{ML} model inputs~\cite{zhang_evaluating_2022, akhtar_advances_2021}, or on real-world scenarios, such as crafting adversarial patterns or objects~\cite{wei_visually_2023, guesmi_physical_2023, eykholt_robust_2018}, they \textit{overlook the impact of different layers of the image processing pipeline on the attack.} 

Many attacks rely on \textit{varied threat models and assumptions, often neglecting the significance of the entire image processing pipeline}. This oversight leads to inconsistent and misleading risk assessments. For example, our systematic analysis indicates that application layer attacks require more sophisticated capabilities than those aiming at the physical world when considering the complete pipeline. This significantly impacts the feasibility of attacks and, consequently, their associated risks. Without a common threat analysis and risk assessment, \textit{we risk making unrealistic evaluations of certain attacks}. To address this issue, our \ac{SoK} systematically analyzes and classifies security-related studies, adapting the ISO 21434 standard~\cite{international_organization_for_standardization_isosae_2021}, intending to establish a consistent and comprehensive threat model that accurately evaluates risks across the entire pipeline.

While many attacks on the image processing pipeline exist, defensive research is significantly underrepresented. By contrast, many studies in the robustness domain have been more comprehensive and have considered the complex image processing pipeline. However, these studies primarily focus on mitigating environmental influences and improving performance benchmarks~\cite{bijelic_benchmarking_2018} rather than addressing intentionally crafted attacks. As a result, \textit{works from the robustness domain are not sufficiently considered to secure the image processing pipeline}. Our \ac{SoK} aims to bridge this critical gap by integrating insights from both security and robustness domains. By doing so, we seek to help researchers develop a more resilient system that combines the strengths of works from both domains.
	
Finally, the complexity of image processing pipelines and their extensive configuration options reveal another significant gap: \textit{no existing testbed in the scientific community effectively addresses the full spectrum of pipeline layers and configurations}. Existing studies often rely on simulations~\cite{blasinski_optimizing_2018, boloor_attacking_2020} or use unrealistic hardware (e.g., smartphones~\cite{cao_invisible_2021} or semi-professional cameras~\cite{yu_reconfigisp_2021}). Furthermore, prior hardware testbeds~\cite{rowe_cmucam3_2007, adams_frankencamera_2010} lack full pipeline coverage and up-to-date technology. Our paper addresses this gap by proposing PICT (\textbf{P}ICT is an \textbf{I}mage \textbf{C}apture \textbf{T}estbed), a new testbed that integrates carefully selected hardware that mimics contemporary automotive systems, covers all pipeline layers, is cost-effective, open-source, and easy to reproduce. PICT provides insights into both the security and robustness of the image processing pipeline and facilitates testing attacks and countermeasures.

We make the following contributions as shown in \autoref{fig:contribution}:
\begin{itemize}
    \item \textbf{Systematizing security and robustness research.} We provide a comprehensive analysis of 92 papers -- 56 on security and 36 on robustness. For each domain, we \textit{classify} and \textit{contextualize} the existing work, assess its impact, and identify key observations (\secref{sec:securityConsiderations} and \secref{sec:robustnessConsiderations}).
    \item \textbf{Evaluating risk and threat models of existing attacks.} To guide future research and enable focused investigation of critical threats, we systematically \textit{analyze} and \textit{classify} security-related research using the ISO 21434 standard (\secref{sec:threatModelConsiderations}). We also release the open-source tool TARA-CAM\footnote{\url{https://tum-esi.github.io/PICT/}}, enabling replication of our analysis and consistent evaluation of similar attacks, thereby offering a more consistent approach to threat assessment.
    \item \textbf{Bridging security and robustness research.} Based on the analysis of research papers from both security and robustness domains, we establish links to increase awareness across the domains for more integrated future research (\secref{sec:linkroubandsec}).
    \item \textbf{Introducing PICT, a testbed for the image processing pipeline.} We developed an open-source\footnote{\url{https://github.com/tum-esi/PICT}} testbed, PICT, deployed on embedded hardware to address the current gap in evaluating image processing pipelines. PICT integrates carefully selected hardware and covers all pipeline layers. It aims to enhance the understanding of attack impact and the effectiveness of countermeasures on image processing pipelines. To demonstrate its usability, we use the tool to showcase representative attacks and the effectiveness of defense mechanisms (\secref{sec:imageSensorTestbed} and \secref{sec:pictevaluation}).
\end{itemize}

\section{Image Processing Pipeline}
\label{sec:detailsImagePipeline}

\autoref{fig:teaser} illustrates a manufacturer-agnostic image processing pipeline, which consists of four main layers: physical world, sensor layer, data preparation layer, and application layer. Data captured from the physical world moves from the sensor layer via the data preparation layer to the application layer. Additionally, control signals from software components on the application layer can influence the data preparation layer and the sensor layer. It is important to note that we use these layers conceptually; thus, the mapping of these layers to actual hardware platforms may vary. For example, components of the data preparation layer and the application layer can run on the same physical device~\cite{jones_6_2024, keall_method_2008}. Similarly, the application layer may be integrated into different hardware components. While real-world automotive image processing pipelines include manufacturer-specific constraints and parameters, they share common functional stages, enabling manufacturer-agnostic analysis.

\cpar{Physical World}
\label{sec:physicalWorld} 
The physical world represents all visible objects in the field of view, such as street signs, road markings, pedestrians, and road users. Since an image sensor is sensitive to photons, it will capture all entities, such as objects or displayed content, emitting or reflecting photons. The more photons an entity emits or passively reflects, the higher the captured and digitized value will be at a dedicated position. Therefore, other influences in the physical world, like sunbeams, shadows, or weather~\cite{ceccarelli_rgb_2023}, can significantly impact the subsequent layers of the image processing pipeline.

\cpar{Sensor Layer}
\label{sec:sensorLayer}
It is the first layer that senses and processes the photons emitted by the physical world. It is important to differentiate between the terms \textit{camera} and \textit{image sensor}. While every camera includes an image sensor, it typically comprises additional optical and electrical components, as shown in \autoref{fig:teaser}. Within this layer, photons are directed through one or more lenses~\cite{florin_simulation_2007}. These lenses can either be fixed in position, providing a fixed focus, or be adjustable, allowing dynamic focus control~\cite{ramanath_color_2005}. Once the light passes through the lenses, it reaches the image sensor.

This sensing element can be a \ac{CCD} or \ac{CMOS} array composed of usually multiple millions of photosensitive elements. Since these photosensitive elements cannot differentiate colors, a \ac{CFA} is placed on top of the sensing array~\cite{el_gamal_cmos_2005}. Various \acp{CFA} are available for different use cases~\cite{ramanath_color_2005}, with the Bayer pattern being the most prominent \ac{CFA}~\cite{florin_simulation_2007}. A Bayer pattern consists of a $2\times 2$ pixel matrix, with one \textbf{r}ed, two \textbf{g}reen, and one \textbf{b}lue  pixel (also called RGGB). Especially in the automotive industry, different variations of this original Bayer pattern exist, such as RCCC or RCCB, where "C" stands for "clear" pixels, meaning that they are not sensitive to only a specific color~\cite{weikl_optimization_2020}. \autoref{fig:teaser} shows different \acp{CFA} which are common in research and industry.

After sensing light via \ac{CCD} or \ac{CMOS} technology, a readout circuit converts the analog measured values into digital ones for further processing~\cite{avidan_all_2022}. For \ac{CMOS} sensors, this readout circuit often operates in a \textit{rolling-shutter} mode, where the entire pixel array is read out line by line~\cite{sayles_invisible_2021} in contrast to \ac{CCD} sensors~\cite{kohler_they_2021}. Once digital values are available, many manufacturer-specific on-chip processing steps are performed~\cite{ceccarelli_rgb_2023, blasinski_optimizing_2018}. These operations can include defect pixel correction, image cropping, tone mapping, and simple denoising~\cite{ramanath_color_2005}. In the automotive industry, \ac{HDR} image sensors are often used, with a higher dynamic range than typical sensors. This is achieved either by combining multiple photosensitive elements per pixel~\cite{willassen_1280x1080_2015} or by merging images of the same scene captured with different settings~\cite{mann_being_1994, heide_flexisp_2014}. Finally, the partially processed digital image is serialized into a data stream for electronic transmission. A typical interface for this purpose is the  \ac{CSI-2} by the MIPI Alliance~\cite{mipi_alliance_camera_2024}. For data transmission over longer distances, the \ac{CSI-2} protocol is often serialized using proprietary standards~\cite{texas_instruments_design_2019}.

\cpar{Data Preparation Layer}
\label{sec:dataPreparationLayer}
The first operation within this layer involves deserializing the received data and delivering it to an \ac{ISP} for further processing. The primary goal of an \ac{ISP} is to optimally prepare the captured image for subsequent use, such as displaying it for the human eye~\cite{ramanath_color_2005} or preparing it for \ac{CV} algorithms~\cite{molloy_impact_2023}. Similar to on-chip processing steps, calculations in the \ac{ISP} are often manufacturer-specific and can include a variety of operations~\cite{arm_ltd_arm_2020}. Therefore, we will highlight the functionality of some typical \ac{ISP} components. The most prominent functionality is demosaicing, which converts the captured raw image format (e.g., Bayer pattern) into an RGB image, where each pixel represents three colors instead of individual color components~\cite{ramanath_color_2005}. Since this spatial operation can introduce artifacts, various demosaicing methods, and subsequent correction techniques are available~\cite{susstrunk_color_2024}. Additionally, the bit depth per pixel is typically reduced, as the raw data from the image sensor often has a high bit depth~\cite{omnivision_technologies_ox03f10_2022}. On the RGB image, many further image quality-enhancing processing steps can be performed, such as color or tone correction~\cite{blasinski_optimizing_2018} or extended image denoising~\cite{baek_noise_2008}. Lastly, \acp{ISP} can perform image scaling and/or cropping steps to prepare the captured image data for the subsequent application~\cite{xiao_seeing_2019}. Additionally, full-size or scaled images can often be lossy compressed by the \ac{ISP}~\cite{ramanath_color_2005}. The prepared image is often used by the application layer to update image sensor settings such as the exposure control or camera gains~\cite{yahiaoui_overview_2019}. Sophisticated image sensors have such functionality already on-chip~\cite{sony_semiconductor_solutions_corporation_isx020_2022}. 

\cpar{Application Layer}
\label{sec:applicationLayer}
The final layer of the image processing pipeline includes various applications for different tasks. For autonomous vehicles, \ac{CV} algorithms are the most important. Among these, \ac{ML} models are the most common for fully autonomous vehicles. These models can range from simple object detection algorithms to complex systems, such as trajectory planning based on images or even end-to-end autonomous driving models~\cite{chen_end--end_2024}. Other applications include the fusion of image data with other sensor data, such as LiDAR,  the visualization of images in the vehicle’s infotainment system, and many others.
In addition, this layer includes applications that control the parameters of preceding components. Examples include configuring the \ac{ISP} or adjusting sensor settings to influence exposure control. This feedback loop is depicted in \autoref{fig:teaser} as a control flow from the application layer to the \ac{ISP} and the image sensor. Similar to the data preparation layer, sophisticated image sensors might have application layer functionality already on-chip~\cite{sony_group_corporation_imx500_2025}.
\section{Security Classification}
\label{sec:securityConsiderations}
After introducing the image processing pipeline, we now analyze the security research related to the various layers. We reviewed 56 security-related research papers with the methodology for their selection detailed in Appendix~\ref{app:researchmethod}. From this analysis, we define \textbf{eight attack classes} to classify the different types of attacks discussed in these papers. Additionally we categorize them into three groups
\begin{enumerate*}[before=\unskip{: }, itemjoin={{; }}, itemjoin*={{, or }}, label={(\roman*)}]
\item Attack Papers~\stackinset{c}{0pt}{c}{1.4ex}{\scalebox{0.7}{\harveyBallFull}}{\scalebox{0.7}{\harveyBallNone}}: Papers that present an attack against one or more of the pipeline components.
This category includes papers that mention possible defenses for the discussed attack but without thorough evaluation~\stackinset{c}{0pt}{c}{1.4ex}{\scalebox{0.7}{\harveyBallFull}}{\scalebox{0.7}{\harveyBallHalf}}
\item Attack and Defense Papers~\stackinset{c}{0pt}{c}{1.4ex}{\scalebox{0.7}{\harveyBallFull}}{\scalebox{0.7}{\harveyBallFull}}: Papers that propose an attack and also include respective defense methods with evaluation
\item Defense Papers~\stackinset{c}{0pt}{c}{1.4ex}{\scalebox{0.7}{\harveyBallNone}}{\scalebox{0.7}{\harveyBallFull}}: Papers that focus solely on defense methods.
\end{enumerate*}
We then map each security-related work to the corresponding layer of the image processing pipeline and highlight the respective vulnerable asset in \autoref{tab:SecurityLayer}.

\begin{table*}[t]
\centering
    \caption{The mapping of the security-related work into the four layers of the image processing pipeline and the eight attack classes: \circledtexted{1} Dynamic Physical Adversarial Distortion, \circledtexted{2} Static Physical Adversarial Distortion, \circledtexted{3} Targeted Light, \circledtexted{4} Invasive Environmental Influences, \circledtexted{5} Image Processing Attacks, \circledtexted{6} Rolling Shutter Attacks, \circledtexted{7} Digital Adversarial Distortion, and \circledtexted{8} Time Delay Attacks. \circledtexted{\small D} refers to pure defensive work.}
    \label{tab:SecurityLayer}
    \setlength{\tabcolsep}{1pt}
    \renewcommand{\arraystretch}{1.5}
    \resizebox{0.95\linewidth}{!}{%
    \begin{tabular}{|*{2}{c}|*{22}{c}|*{18}{c}|*{6}{c}|*{10}{c}|}\hline
    \multicolumn{2}{|c|}{\textbf{Layer}} & \multicolumn{22}{c|}{\textbf{Physical}} & \multicolumn{18}{c|}{\textbf{Sensor}} & \multicolumn{6}{c|}{\textbf{Data Prep.}} & \multicolumn{10}{c|}{\textbf{Application}} \\\hline
    \multicolumn{2}{|c|}{\textbf{Work}} & \rotatebox[origin=c]{90}{\cite{patel_overriding_2022}} & \rotatebox[origin=c]{90}{\cite{gnanasambandam_optical_2021}} & \rotatebox[origin=c]{90}{\cite{nassi_phantom_2020}} & \rotatebox[origin=c]{90}{\cite{lovisotto_slap_2021}} & \rotatebox[origin=c]{90}{\cite{athalye_synthesizing_2018}} & \rotatebox[origin=c]{90}{\cite{eykholt_robust_2018}} & \rotatebox[origin=c]{90}{\cite{radutoiu_physical_2023}} & \rotatebox[origin=c]{90}{\cite{jan_connecting_2019}} & \rotatebox[origin=c]{90}{\cite{kong_physgan_2020}} & \rotatebox[origin=c]{90}{\cite{wang_dual_2021}} & \rotatebox[origin=c]{90}{\cite{huang_universal_2020}} & \rotatebox[origin=c]{90}{\cite{he_dorpatch_2024}} & \rotatebox[origin=c]{90}{\cite{jing_too_2021}} & \rotatebox[origin=c]{90}{\cite{jia_fooling_2022}} & \rotatebox[origin=c]{90}{\cite{zhao_seeing_2019}} & \rotatebox[origin=c]{90}{\cite{mohajeransari_discovering_2024}} & \rotatebox[origin=c]{90}{\cite{wang_rfla_2023}} & \rotatebox[origin=c]{90}{\cite{petit_remote_2015}} & \rotatebox[origin=c]{90}{\cite{bhupathiraju_vulnerability_2024}} & \rotatebox[origin=c]{90}{\cite{duan_adversarial_2021}} & \rotatebox[origin=c]{90}{\cite{hsiao_natural_2024}} & \rotatebox[origin=c]{90}{\cite{feng_fight_2025}} & \rotatebox[origin=c]{90}{\cite{xia_moire_2024}} & \rotatebox[origin=c]{90}{\cite{man_remote_2024}} & \rotatebox[origin=c]{90}{\cite{zhou_doublestar_2022}} & \rotatebox[origin=c]{90}{\cite{sato_invisible_2024}} & \rotatebox[origin=c]{90}{\cite{wang_i_2021}} & \rotatebox[origin=c]{90}{\cite{ji_poltergeist_2021}} & \rotatebox[origin=c]{90}{\cite{kohler_signal_2022}} & \rotatebox[origin=c]{90}{\cite{long_side_2023}} & \rotatebox[origin=c]{90}{\cite{oyama_adversarial_2024}} & \rotatebox[origin=c]{90}{\cite{huang_lights_2022}} & \rotatebox[origin=c]{90}{\cite{wu_illumination_2021}} & \rotatebox[origin=c]{90}{\cite{sayles_invisible_2021}} & \rotatebox[origin=c]{90}{\cite{guo_invisible_2024}} & \rotatebox[origin=c]{90}{\cite{li_light_2020}} & \rotatebox[origin=c]{90}{\cite{yan_rolling_2022}} & \rotatebox[origin=c]{90}{\cite{kohler_they_2021}} & \rotatebox[origin=c]{90}{\cite{liu_cross-task_2024}} & \rotatebox[origin=c]{90}{\cite{mipi_alliance_guide_2024}} & \rotatebox[origin=c]{90}{\cite{phan_adversarial_2021}} & \rotatebox[origin=c]{90}{\cite{li_adversarial_2022}} & \rotatebox[origin=c]{90}{\cite{li_image-scaling_2024}} & \rotatebox[origin=c]{90}{\cite{xiao_seeing_2019}} & \rotatebox[origin=c]{90}{\cite{quiring_adversarial_2020}} & \rotatebox[origin=c]{90}{\cite{avidan_all_2022}} & \rotatebox[origin=c]{90}{\cite{agarwal_crafting_2022}} & \rotatebox[origin=c]{90}{\cite{xiong_multi-source_2021}} & \rotatebox[origin=c]{90}{\cite{boloor_attacking_2020}} & \rotatebox[origin=c]{90}{\cite{sato_robustness_2021}} & \rotatebox[origin=c]{90}{\cite{ma_slowtrack_2024}} & \rotatebox[origin=c]{90}{\cite{muller_investigating_2025}} & \rotatebox[origin=c]{90}{\cite{gurel_knowledge_2021}} & \rotatebox[origin=c]{90}{\cite{man_that_2023}} & \rotatebox[origin=c]{90}{\cite{zhang_detecting_2021}} & \rotatebox[origin=c]{90}{\cite{luo_foveation-based_2016}} \\ \hline
    \multicolumn{2}{|c|}{\textbf{Class}} & \circledtexted{1} & \circledtexted{1} & \circledtexted{1} & \circledtexted{1} & \circledtexted{2} & \circledtexted{2} & \circledtexted{2} & \circledtexted{2} & \circledtexted{2} & \circledtexted{2} & \circledtexted{2} & \circledtexted{2} & \circledtexted{2} & \circledtexted{2} & \circledtexted{2} & \circledtexted{2} & \circledtexted{3} & \circledtexted{3} & \circledtexted{3} & \circledtexted{3} & \circledtexted{3} & \circledtexted{\small D} & \circledtexted{1} & \circledtexted{1} & \circledtexted{3} & \circledtexted{3} & \circledtexted{3} & \circledtexted{4} & \circledtexted{4} & \circledtexted{4} & \circledtexted{5} & \circledtexted{6} & \circledtexted{6} & \circledtexted{6} & \circledtexted{6} & \circledtexted{6} & \circledtexted{6} & \circledtexted{6} & \circledtexted{6} & \circledtexted{\small D} & \circledtexted{5} & \circledtexted{5} & \circledtexted{5} & \circledtexted{5} & \circledtexted{\small D} & \circledtexted{\small D} & \circledtexted{5} & \circledtexted{7} & \circledtexted{7} & \circledtexted{7} & \circledtexted{8} & \circledtexted{8} & \circledtexted{\small D} & \circledtexted{\small D} & \circledtexted{\small D} & \circledtexted{\small D} \\ \hline
    \multicolumn{2}{|c|}{\textbf{Vuln.\,Asset}} & \rotatebox[origin=c]{90}{Perception} & \rotatebox[origin=c]{90}{Perception} & \rotatebox[origin=c]{90}{Planning} & \rotatebox[origin=c]{90}{Perception} & \rotatebox[origin=c]{90}{Perception} & \rotatebox[origin=c]{90}{Perception} & \rotatebox[origin=c]{90}{Fusion} & \rotatebox[origin=c]{90}{Perception} & \rotatebox[origin=c]{90}{Perception} & \rotatebox[origin=c]{90}{Perception} & \rotatebox[origin=c]{90}{Perception} & \rotatebox[origin=c]{90}{Perception} & \rotatebox[origin=c]{90}{Perception} & \rotatebox[origin=c]{90}{Perception} & \rotatebox[origin=c]{90}{Perception} & \rotatebox[origin=c]{90}{Planning} & \rotatebox[origin=c]{90}{Perception} & \rotatebox[origin=c]{90}{\ac{CMOS}/\ac{CCD}} & \rotatebox[origin=c]{90}{Perception} & \rotatebox[origin=c]{90}{Perception} & \rotatebox[origin=c]{90}{Perception} & \rotatebox[origin=c]{90}{--} & \rotatebox[origin=c]{90}{\ac{CMOS}/\ac{CCD}} & \rotatebox[origin=c]{90}{Lens} & \rotatebox[origin=c]{90}{\ac{CMOS}/\ac{CCD}} & \rotatebox[origin=c]{90}{\ac{CMOS}/\ac{CCD}} & \rotatebox[origin=c]{90}{\ac{CMOS}/\ac{CCD}} & \rotatebox[origin=c]{90}{Lens} & \rotatebox[origin=c]{90}{\ac{CCD}} & \rotatebox[origin=c]{90}{Lens} & \rotatebox[origin=c]{90}{Ser./Deser.} & \rotatebox[origin=c]{90}{\ac{CMOS}} & \rotatebox[origin=c]{90}{\ac{CMOS}} & \rotatebox[origin=c]{90}{\ac{CMOS}} & \rotatebox[origin=c]{90}{\ac{CMOS}} & \rotatebox[origin=c]{90}{\ac{CMOS}} & \rotatebox[origin=c]{90}{\ac{CMOS}} & \rotatebox[origin=c]{90}{\ac{CMOS}} & \rotatebox[origin=c]{90}{\ac{CMOS}} & \rotatebox[origin=c]{90}{--} & \rotatebox[origin=c]{90}{\ac{ISP}} & \rotatebox[origin=c]{90}{\ac{ISP}} & \rotatebox[origin=c]{90}{\ac{ISP}} & \rotatebox[origin=c]{90}{\ac{ISP}} & \rotatebox[origin=c]{90}{--} & \rotatebox[origin=c]{90}{--} & \rotatebox[origin=c]{90}{Perception} & \rotatebox[origin=c]{90}{Fusion} & \rotatebox[origin=c]{90}{Control} & \rotatebox[origin=c]{90}{Perception} & \rotatebox[origin=c]{90}{Perception} & \rotatebox[origin=c]{90}{Perception} & \rotatebox[origin=c]{90}{--} & \rotatebox[origin=c]{90}{--} & \rotatebox[origin=c]{90}{--} & \rotatebox[origin=c]{90}{--} \\ \hline
    \multirow{2}{*}{\adjustbox{valign=c,rotate=90, scale=0.8}{\textbf{Category}}}  & \multicolumn{1}{|c|}{\textbf{Attack}} & \harveyBallFull & \harveyBallFull & \harveyBallFull & \harveyBallFull & \harveyBallFull & \harveyBallFull & \harveyBallFull & \harveyBallFull & \harveyBallFull & \harveyBallFull & \harveyBallFull & \harveyBallFull & \harveyBallFull & \harveyBallFull & \harveyBallFull & \harveyBallFull & \harveyBallFull & \harveyBallFull & \harveyBallFull & \harveyBallFull & \harveyBallFull & \harveyBallNone & \harveyBallFull & \harveyBallFull & \harveyBallFull & \harveyBallFull & \harveyBallFull & \harveyBallFull & \harveyBallFull & \harveyBallFull & \harveyBallFull & \harveyBallFull & \harveyBallFull & \harveyBallFull & \harveyBallFull & \harveyBallFull & \harveyBallFull & \harveyBallFull & \harveyBallFull & \harveyBallNone & \harveyBallFull & \harveyBallFull & \harveyBallFull & \harveyBallFull & \harveyBallNone & \harveyBallNone & \harveyBallFull & \harveyBallFull & \harveyBallFull & \harveyBallFull & \harveyBallFull & \harveyBallFull & \harveyBallNone & \harveyBallNone & \harveyBallNone & \harveyBallNone \\
    & \multicolumn{1}{|c|}{\textbf{Defense}} & \harveyBallNone & \harveyBallNone & \harveyBallFull & \harveyBallFull & \harveyBallNone & \harveyBallNone & \harveyBallNone & \harveyBallNone & \harveyBallNone & \harveyBallNone & \harveyBallNone & \harveyBallHalf & \harveyBallHalf & \harveyBallHalf & \harveyBallFull & \harveyBallFull & \harveyBallNone & \harveyBallHalf & \harveyBallHalf & \harveyBallFull & \harveyBallFull & \harveyBallFull & \harveyBallFull & \harveyBallFull & \harveyBallHalf & \harveyBallFull & \harveyBallFull & \harveyBallNone & \harveyBallHalf & \harveyBallFull & \harveyBallNone & \harveyBallNone & \harveyBallNone & \harveyBallNone & \harveyBallHalf & \harveyBallFull & \harveyBallFull & \harveyBallFull & \harveyBallFull & \harveyBallFull & \harveyBallNone & \harveyBallHalf & \harveyBallHalf & \harveyBallFull & \harveyBallFull & \harveyBallFull & \harveyBallFull & \harveyBallNone & \harveyBallNone & \harveyBallNone & \harveyBallNone & \harveyBallFull & \harveyBallFull & \harveyBallFull & \harveyBallFull & \harveyBallFull \\ \hline
    \multicolumn{30}{l}{\harveyBallFull: Yes, \harveyBallNone: No, \harveyBallHalf: Possible defenses without further evaluation}
    \end{tabular}
    }
\end{table*}

\subsection{Physical World}

In the physical world, attacks are separated into three classes: \circledtexted{1}~\textit{Dynamic Physical Adversarial Distortions}, \circledtexted{2}~\textit{Static Physical Adversarial Distortions}, and \circledtexted{3}~\textit{Targeted Light Attacks}. Additionally, it contains \circledtexted{\small D}~\textit{Defenses}.
Many existing surveys summarize the research on these physical adversarial samples~\cite{akhtar_advances_2021, guesmi_physical_2023, wei_visually_2023, wei_physical_2024, wang_survey_2023, wang_does_2023}. We emphasize that more sophisticated physical attacks, exploiting sensor characteristics~\cite{man_remote_2024} are covered separately in~\secref{sec:securityWorkSensor}.

\cpar{\circledtexted{1}~Dynamic Physical Adversarial Distortions}
Attackers can project~\cite{nassi_phantom_2020, gnanasambandam_optical_2021, lovisotto_slap_2021} or display images~\cite{patel_overriding_2022} on physical layer items to disrupt the functionality of cameras in autonomous vehicles. While~\cite{patel_overriding_2022, gnanasambandam_optical_2021} solely show attacks, \cite{nassi_phantom_2020, lovisotto_slap_2021} propose additional mitigation strategies via adversarial training~\cite{lovisotto_slap_2021} or improved ML models~\cite{nassi_phantom_2020}. All these attacks can be adapted dynamically to change, for example, the size or the content of the adversarial distortions. While this typically increases the attack complexity, it can compensate for changing environmental conditions or viewing angles.

\cpar{\circledtexted{2}~Static Physical Adversarial Distortions}
By using printed images~\cite{jan_connecting_2019, kong_physgan_2020, eykholt_robust_2018, jia_fooling_2022, wang_dual_2021, huang_universal_2020}, stickers~\cite{zhao_seeing_2019, eykholt_robust_2018, he_dorpatch_2024, jing_too_2021}, or even three dimensional objects~\cite{athalye_synthesizing_2018, radutoiu_physical_2023}, attackers can create persistent attacks in the physical world. Typically, environmental influences must be considered when creating such static distortions~\cite{jan_connecting_2019, kong_physgan_2020, zhao_seeing_2019, jia_fooling_2022}. Additionally, such perturbations could be overlays on complete objects~\cite{wang_dual_2021, huang_universal_2020} or stickers only on a small region~\cite{eykholt_robust_2018}. While most work~\cite{athalye_synthesizing_2018, eykholt_robust_2018, radutoiu_physical_2023, jan_connecting_2019, kong_physgan_2020, jia_fooling_2022, he_dorpatch_2024, wang_dual_2021} provide attacks, some also provide mitigations~\cite{zhao_seeing_2019, mohajeransari_discovering_2024}. Unlike \circledtexted{1} dynamic physical adversarial distortions, static ones remain fixed and cannot be changed during deployment since they need to be recreated or newly placed.

\cpar{\circledtexted{3}~Targeted Light Attacks}
With targeted light, attackers can either directly blind cameras~\cite{petit_remote_2015} and saturate the resulting image or use reflections from natural~\cite{hsiao_natural_2024, wang_rfla_2023} or artificial light sources~\cite{bhupathiraju_vulnerability_2024, duan_adversarial_2021}. In most cases, laser pointers, creating a strong light beam, are typical attack devices~\cite{petit_remote_2015, bhupathiraju_vulnerability_2024, duan_adversarial_2021}. These non-persistent attacks can also be found in other application domains of cameras~\cite{fu_remote_2022, hutchison_preventing_2005}.

\cpar{\circledtexted{\small D}~Defenses}
\citeauthor{feng_fight_2025}~\cite{feng_fight_2025} introduce static physical adversarial distortions that can be used to detect and defend against malicious physical adversarial distortions.

\subsection{Sensor Layer}
\label{sec:securityWorkSensor}

The sensor layer is critical because it is the first point where researchers and engineers can actively influence system design, offering opportunities not only for attacks but also for defending the image processing pipeline. There is a variety of available attack classes: \circledtexted{1}~\textit{Dynamic Physical Adversarial Distortions}, \circledtexted{3}~\textit{Targeted Light Attacks}, \circledtexted{4}~\textit{Invasive Environmental Influences}, \circledtexted{5}~\textit{Image Processing Attacks}, \circledtexted{6}~\textit{Rolling Shutter Attacks}, and a \circledtexted{\small D} Defense.

\cpar{\circledtexted{1}~Dynamic Physical Adversarial Distortions}
By using projectors, attackers can exploit lens flare effects, causing optical artifacts due to reflections~\cite{man_remote_2024}. Another attack, presented by \citeauthor{xia_moire_2024}~\cite{xia_moire_2024}, exploits the array structure of photosensitive elements in image sensors when showing images on a display. They demonstrate that the Moir\'{e} effect, typically seen when geometric patterns such as lines intersect at certain angles~\cite{oster_moire_1963}, can occur on image sensors, adversely affecting object classification.

\cpar{\circledtexted{3}~Targeted Light Attacks}
Controlled targeted light can be used to exploit specific characteristics of image sensors, such as their sensitivity to infrared light~\cite{sato_invisible_2024, wang_i_2021} or specific lenses~\cite{zhou_doublestar_2022}. Although these attacks are comparable to \circledtexted{3}~targeted light attacks in the physical world, they have sensor dependencies and are therefore categorized in the sensor layer.

\cpar{\circledtexted{4}~Invasive Environmental Influences}
In contrast to the previous attacks, invasive environmental influences aim to attack side-channels of image sensors by introducing acoustic~\cite{ji_poltergeist_2021, long_side_2023} or electromagnetic interference~\cite{kohler_signal_2022}. However, these attacks show limitations in their applicability: While electromagnetic interference works only on \ac{CCD}-based image sensors~\cite{kohler_signal_2022}, existing acoustic attacks~\cite{ji_poltergeist_2021, long_side_2023} work only on cameras with movable lens systems.

\cpar{\circledtexted{5}~Image Processing Attacks}
\citeauthor{oyama_adversarial_2024}~\cite{oyama_adversarial_2024} explore an attack targeting the physical connection between the camera and its subsequent device, injecting manipulated image data. It is placed at the boundary between the sensor and data preparation layer.

\cpar{\circledtexted{6}~Rolling Shutter Attacks}
As \ac{CMOS}-based image sensors are the most common image sensor technology in autonomous vehicles, their rolling shutter effect represents a significant vulnerability. Rolling shutter attacks~\cite{sayles_invisible_2021, liu_cross-task_2024, yan_rolling_2022, guo_invisible_2024} represent illumination-based attacks, working with pulsed light at a high frequency. Similar attacks~\cite{huang_lights_2022, wu_illumination_2021, li_light_2020} can also disturb face detection functionality that can be relevant for in-vehicle monitoring systems.

\cpar{\circledtexted{\small D}~Defenses}
The MIPI Alliance introduced a security framework, enabling encryption and authentication already on the image sensor, to prevent further tampering with the digitized image~\cite{mipi_alliance_guide_2024}.

\observation{System design decisions of image sensors, such as the rolling-shutter mechanism, can lead to security vulnerabilities. Such design flaws can propagate to software applications that are often not secured against these attacks.}

\subsection{Data Preparation Layer}
Within this layer, different image processing steps enhance image quality and prepare data for \ac{CV} algorithms. However, some of these operations are susceptible to attacks. We classify these as \circledtexted{5}~\textit{Image Processing Attacks} and their respective \circledtexted{\small D}~\textit{Defense}.

\cpar{\circledtexted{5}~Image Processing Attacks}
By exploiting \ac{ISP} characteristics, it is possible to create camera-specific attacks that hide information tailored to work only on specific \acp{ISP}~\cite{phan_adversarial_2021}. Additionally, certain processing steps of an \ac{ISP} can be exploited in general, such as image scaling~\cite{li_adversarial_2022, li_image-scaling_2024, xiao_seeing_2019} by embedding crafted information that only becomes visible after scaling the image, leading to misclassification in subsequent layers. Through low-level image analysis, this attack can be thwarted~\cite{xiao_seeing_2019}.

\cpar{\circledtexted{\small D} Defenses}
It is possible to mitigate the effect of general artificial perturbations by processing the raw image through different \ac{ISP} pipelines~\cite{avidan_all_2022}. Additionally, improved scaling algorithms are available that can mitigate the risk of image-scaling attacks~\cite{quiring_adversarial_2020}.

\observation{Processing steps in the data preparation layer can significantly alter the appearance of images, such as the color, spatial artifacts, or image size. Attackers can exploit these processing steps to craft attacks, to create very targeted, asset-specific attacks~\cite{phan_adversarial_2021}. Still, as \autoref{tab:comparisonSecurity} shows, the amount of security-related research is low compared to other layers, suggesting an opportunity to study ISP-based threats and defenses.}

\subsection{Application Layer}
\label{sec:securityWorkApplication}

Given the flexibility and variety of options within the application layer, we focus on work specifically related to autonomous vehicles or realistic use cases where images are directly evaluated. Attacks and defenses requiring non-temporary image storage or extensive post-processing are excluded, as these methods are not applicable to fully autonomous vehicles with onboard \ac{CV} algorithms. In this layer, three different attack classes can be observed: \circledtexted{5}~\textit{Image Processing Attacks}, \circledtexted{7}~\textit{Digital Adversarial Distortions}, and \circledtexted{8}~\textit{Time Delay Attacks}. Additionally, there are multiple \circledtexted{\small D}~\textit{Defenses} available.

\cpar{\circledtexted{5}~Image Processing Attacks}
Since the application layer works on fully processed images, it allows adversarial perturbations via modifications of image transformation parameters in such prepared images~\cite{agarwal_crafting_2022} through further image processing. This impact can be mitigated using techniques such as adversarial training~\cite{agarwal_crafting_2022}, which fall outside the scope of this \ac{SoK}’s focus.

\cpar{\circledtexted{7}~Digital Adversarial Distortions}
Digital adversarial distortions can impact new autonomous driving paradigms like end-to-end autonomous driving~\cite{boloor_attacking_2020}, or even multi-sensor systems~\cite{xiong_multi-source_2021}, consisting of cameras and LiDARs. Additionally, specific automotive perception algorithms such as lane detection~\cite{sato_robustness_2021} can be attacked.

\cpar{\circledtexted{8}~Time Delay Attacks}
By introducing specific patterns in images, the execution time of object detection algorithms can be increased and disturb the system functionality~\cite{ma_slowtrack_2024, muller_investigating_2025}. Although \citeauthor{muller_investigating_2025}~\cite{muller_investigating_2025} propose patches that are placed in the physical world, they attack specific object detection algorithms in the application layer as vulnerable assets.

\cpar{\circledtexted{\small D}~Defenses}
Many defense strategies focus on optimizing \ac{ML} models or adjusting their input parameters~\cite{akhtar_advances_2021, kyrkou_towards_2020, akhtar_defense_2018, zhang_defense_2021, nie_diffusion_2022}. However, these approaches often overlook the earlier layers in the overall pipeline. As part of this \ac{SoK}, we do not focus on well-known defense \ac{ML} methods like reliable adversarial training~\cite{madry_towards_2018} or data randomization during training~\cite{xie_mitigating_2018}. Instead, we concentrate on work that considers elements of the image processing pipeline. Some existing research considers countermeasures against attacks on the components shown in \autoref{fig:teaser}. For example, there is work on detecting physical camera blinding attacks~\cite{zhang_detecting_2021}. Other studies focus on regions of interest~\cite{luo_foveation-based_2016}, use multiple \ac{ML} models for knowledge enhancement~\cite{gurel_knowledge_2021}, or include consistency checks with other available information in autonomous vehicles~\cite{man_that_2023}.
\section{Risk \& Threat Model Classification}
\label{sec:threatModelConsiderations}

\begin{table*}[t!]
\centering
  \caption{Evaluation of attacks sorted by layer and evaluated risk. Category details are available in Appendix~\ref{app:riskAnalysisCameraAttacks}.}
  \label{tab:comparisonSecurity}
  \begin{adjustbox}{totalheight=.96\textheight-2\baselineskip}
  \begin{tabular}{|c|c|c|ccc|c|cccc|c|c|}
  \hline
\rowcolor{gray!5}  \textbf{\rotatebox{90}{Layer}} & \textbf{\rotatebox{90}{Work}} &  \textbf{\rotatebox{0}{Entry Point}} & \textbf{\rotatebox{90}{Impact S}} & \textbf{\rotatebox{90}{Impact O}} & \textbf{\rotatebox{90}{Accuracy}} & \textbf{\rotatebox{90}{Impact}}& \textbf{\rotatebox{90}{Knowledge}} & \textbf{\rotatebox{90}{Equipment}} & \textbf{\rotatebox{90}{WoO}} & \textbf{\rotatebox{90}{Expertise}} & \textbf{\rotatebox{90}{Feasibility}} & \textbf{\rotatebox{90}{Risk}} \\
  \hline
\multirow{21}{*}{\rotatebox[origin=c]{90}{Physical}}&\cite{patel_overriding_2022}  & Physical World & \impact{4} & \impact{3} & \targeted & \impact{4}  & \WhiteCircle{} & Multiple Bespoke & < 100m & \mexpert & \VeryLowFeasibility{} & \ApplyGradient{1} \\
&\cite{athalye_synthesizing_2018}  & Physical World & \impact{3} & \impact{2} & \targeted  & \impact{3}  & \WhiteCircle{} & Specialized & < 1m & \expert & \VeryLowFeasibility{} & \ApplyGradient{1} \\
&\cite{jing_too_2021}  & Physical World & \impact{2} & \impact{1} & \targeted  & \impact{2}  & \WhiteCircle{} & Multiple Bespoke & < 100m & \mexpert & \VeryLowFeasibility{} & \ApplyGradient{1} \\
&\cite{he_dorpatch_2024}  & Physical World & \impact{2} & \impact{4} & \targeted  & \impact{4}  & \WhiteCircle{} & Specialized & < 1m & \expert & \VeryLowFeasibility{} & \ApplyGradient{1} \\
&\cite{zhao_seeing_2019} & Physical World & \impact{3} & \impact{2} & \targeted  & \impact{3}  & \WhiteCircle{} & Specialized & < 100m & \expert & \LowFeasibility{} & \ApplyGradient{2} \\
&\cite{eykholt_robust_2018} & Physical World & \impact{3} & \impact{2} & \targeted  & \impact{3}  & \WhiteCircle{} & Standard & < 10m & \mexpert & \LowFeasibility{} & \ApplyGradient{2} \\
&\cite{huang_universal_2020} & Physical World & \impact{2} & \impact{3} & \targeted  & \impact{3}  & \WhiteCircle{} & Specialized & < 10m & \expert & \LowFeasibility{} & \ApplyGradient{2} \\
&\cite{radutoiu_physical_2023}  & Physical World & \impact{3} & \impact{2} & \targeted  & \impact{3}  & \WhiteCircle{} & Specialized & < 10m & \expert & \LowFeasibility{} & \ApplyGradient{2}\\
&\cite{jan_connecting_2019} & Physical World & \impact{3} & \impact{2} & \targeted  & \impact{3}  & \GrayCircle{} & Specialized & < 1m & \expert &  \MediumFeasibility{}  & \ApplyGradient{3} \\
&\cite{wang_rfla_2023} & Physical World & \impact{3} & \impact{2} & \ntargeted  & \impact{2}  & \BlackCircle{} & Standard & < 10m & \Layman &  \HighFeasibility{}  & \ApplyGradient{3} \\
&\cite{kong_physgan_2020}  & Physical World & \impact{4} & \impact{2} & \targeted  & \impact{4}  & \WhiteCircle{} & Specialized & < 100m & \expert & \LowFeasibility{} & \ApplyGradient{3} \\
&\cite{wang_dual_2021}  & Physical World & \impact{3} & \impact{3} & \targeted  & \impact{4}  & \BlackCircle{} & Bespoke & < 10m & \expert & \MediumFeasibility{} & \ApplyGradient{4} \\
&\cite{jia_fooling_2022}  & Physical World & \impact{2} & \impact{3} & \targeted  & \impact{3}  & \BlackCircle{} & Specialized & < 100m & \mexpert & \HighFeasibility{} & \ApplyGradient{4} \\
&\cite{petit_remote_2015}  & Physical World & \impact{3} & \impact{3} & \ntargeted  & \impact{3}  & \BlackCircle{} & Standard & < 10m & \Layman & \HighFeasibility{} & \ApplyGradient{4} \\
&\cite{duan_adversarial_2021}  & Physical World & \impact{3} & \impact{3} & \ntargeted  & \impact{3}  & \BlackCircle{} & Standard & < 10m & \prof & \HighFeasibility{} & \ApplyGradient{4} \\
&\cite{hsiao_natural_2024} & Physical World & \impact{3} & \impact{3} & \ntargeted & \impact{3} & \BlackCircle{} & Standard & < 10m & \Layman & \HighFeasibility{} & \ApplyGradient{4} \\
&\cite{mohajeransari_discovering_2024} & Physical World & \impact{4} & \impact{3} & \targeted & \impact{4} & \BlackCircle{} & Multiple Bespoke & < 100m & \expert & \MediumFeasibility{} & \ApplyGradient{4} \\
&\cite{gnanasambandam_optical_2021} & Physical World & \impact{3} & \impact{3} & \targeted  & \impact{4}  & \BlackCircle{} & Specialized & < 10m & \expert & \HighFeasibility{} & \ApplyGradient{5} \\
&\cite{lovisotto_slap_2021} & Physical World & \impact{3} & \impact{3} & \targeted  & \impact{4}  & \BlackCircle{} & Specialized & < 100m & \expert & \HighFeasibility{} & \ApplyGradient{5} \\
&\cite{bhupathiraju_vulnerability_2024} & Physical World & \impact{3} & \impact{4} & \targeted  & \impact{4}  & \BlackCircle{} & Specialized & < 100m & \prof & \HighFeasibility{} & \ApplyGradient{5} \\
&\cite{nassi_phantom_2020} & Physical World & \impact{4} & \impact{3} & \targeted  & \impact{4}  & \BlackCircle{} & Specialized & < 10m & \expert & \HighFeasibility{} & \ApplyGradient{5} \\ 
\hline
\multirow{17}{*}{\rotatebox[origin=c]{90}{Sensor}}&\cite{ji_poltergeist_2021}  & Physical World & \impact{2} & \impact{3} & \ntargeted  & \impact{2}  & \GrayCircle{} & Bespoke & < 0.1m & \mexpert & \VeryLowFeasibility{} & \ApplyGradient{1} \\
&\cite{long_side_2023}  & Physical World & \impact{1} & \impact{3} & \ntargeted  & \impact{2}  & \GrayCircle{}& Specialized & < 0.1m & \expert & \VeryLowFeasibility{} & \ApplyGradient{1} \\
&\cite{oyama_adversarial_2024}  & Sensor / Data & \impact{3} & \impact{2} & \targeted  & \impact{3}  & \WhiteCircle{} & Multiple Bespoke & < 0.1m & \mexpert & \VeryLowFeasibility{} & \ApplyGradient{1}\\
&\cite{li_light_2020} & Physical World & \impact{1} & \impact{2} & \targeted  & \impact{2}  & \BlackCircle{} & Bespoke & < 1m & \prof &  \MediumFeasibility{}  & \ApplyGradient{2}\\
&\cite{zhou_doublestar_2022}  & Physical World  & \impact{3} & \impact{3} & \ntargeted  & \impact{3}  & \BlackCircle{} & Bespoke & < 10m & \expert & \MediumFeasibility{} & \ApplyGradient{3}  \\
&\cite{huang_lights_2022}& Physical World & \impact{1} & \impact{2} & \targeted  & \impact{2}  & \BlackCircle{} & Bespoke & < 10m & \prof & \HighFeasibility{} & \ApplyGradient{3} \\
&\cite{wu_illumination_2021} & Physical World & \impact{1} & \impact{2} & \targeted  & \impact{2}  & \BlackCircle{} & Bespoke & < 10m & \prof & \HighFeasibility{} & \ApplyGradient{3} \\
&\cite{man_remote_2024}  & Physical World & \impact{3} & \impact{2} & \targeted  & \impact{3}  & \GrayCircle{} & Specialized & < 1m & \prof &  \MediumFeasibility{}  &\ApplyGradient{3}  \\
&\cite{kohler_signal_2022}  & Physical World  & \impact{3} & \impact{3} & \targeted  & \impact{4}  & \GrayCircle{} & Bespoke & < 1m & \mexpert & \LowFeasibility{} & \ApplyGradient{3}  \\
&\cite{sato_invisible_2024}  & Physical World & \impact{3} & \impact{3} & \ntargeted  & \impact{3}  & \BlackCircle{} & Bespoke & < 100m & \expert &  \HighFeasibility{}  & \ApplyGradient{4}  \\
&\cite{sayles_invisible_2021}  & Physical World & \impact{3} & \impact{3} & \targeted  & \impact{4}  & \GrayCircle{} & Bespoke & < 0.5m & \prof &  \MediumFeasibility{}  & \ApplyGradient{4}  \\
&\cite{yan_rolling_2022}  & Physical World & \impact{4} & \impact{3} & \targeted  & \impact{4}  & \GrayCircle{} & Bespoke & < 10m & \expert &  \MediumFeasibility{}  & \ApplyGradient{4} \\
&\cite{kohler_they_2021}  & Physical World & \impact{3} & \impact{3} & \ntargeted  & \impact{3}  & \BlackCircle{} & Specialized & < 10m & \expert & \HighFeasibility{} & \ApplyGradient{4} \\
&\cite{liu_cross-task_2024} & Physical World & \impact{3} & \impact{3} & \ntargeted  & \impact{3}  & \BlackCircle{} & Bespoke & < 10m & \prof & \HighFeasibility{} & \ApplyGradient{4} \\
&\cite{guo_invisible_2024} & Physical World & \impact{3} & \impact{3} & \targeted  & \impact{4}  & \BlackCircle{}& Specialized & < 10m & \expert & \HighFeasibility{} & \ApplyGradient{5} \\
&\cite{xia_moire_2024} & Physical World & \impact{3} & \impact{3} & \targeted  & \impact{3}  & \BlackCircle{}& Standard & < 10m & \expert & \HighFeasibility{} & \ApplyGradient{5} \\
&\cite{wang_i_2021}  & Physical World & \impact{4} & \impact{3} & \targeted  & \impact{4}  & \BlackCircle{} & Specialized & < 10m & \expert & \HighFeasibility{} & \ApplyGradient{5} \\ 
\hline
\multirow{5}{*}{\rotatebox[origin=c]{90}{Data Pr.}}&\cite{phan_adversarial_2021} & Physical World & \impact{2} & \impact{3} & \targeted  & \impact{3}  & \BlackCircle{} & Specialized & < 1m & \mexpert &  \MediumFeasibility{}  & \ApplyGradient{3} \\
&\cite{li_image-scaling_2024}  & Physical World & \impact{2} & \impact{3} & \targeted  & \impact{3}  & \BlackCircle{} & Specialized & < 1m & \expert &  \MediumFeasibility{}  & \ApplyGradient{3} \\
&\cite{li_adversarial_2022}  & Application Layer & \impact{3} & \impact{2} & \targeted  & \impact{3}  & \BlackCircle{} & Standard & remote & \expert &  \MediumFeasibility{}  & \ApplyGradient{3} \\
&\cite{xiao_seeing_2019}  & Application Layer & \impact{3} & \impact{2} & \targeted  & \impact{3}  & \BlackCircle{} & Standard & remote & \expert &  \MediumFeasibility{}  & \ApplyGradient{3} \\ 
\hline  
\multirow{6}{*}{\rotatebox[origin=c]{90}{Application}}&\cite{xiong_multi-source_2021}  & Application Layer & \impact{3} & \impact{3} & \targeted  & \impact{4}  & \WhiteCircle{} & Standard & remote & \expert & \VeryLowFeasibility{} & \ApplyGradient{1}\\
& \cite{ma_slowtrack_2024} & Application Layer & \impact{4} & \impact{3} & \ntargeted  & \impact{3}  & \WhiteCircle{} & Standard & remote & \expert &\VeryLowFeasibility{}& \ApplyGradient{1}\\
& \cite{muller_investigating_2025} & Physical World & \impact{3} & \impact{2} & \ntargeted  & \impact{2}  & \WhiteCircle{} & Standard & < 10m & \expert &\MediumFeasibility{}& \ApplyGradient{2}\\
& \cite{sato_robustness_2021} & Physical World & \impact{3} & \impact{3} & \ntargeted  & \impact{3}  & \BlackCircle{} & Bespoke & < 10m & \expert &  \MediumFeasibility{}  & \ApplyGradient{3}\\
& \cite{boloor_attacking_2020}  & Physical World & \impact{4} & \impact{3} & \targeted  & \impact{4}  & \GrayCircle{} & Specialized & < 1m & \expert &  \MediumFeasibility{}  & \ApplyGradient{4} \\
& \cite{agarwal_crafting_2022}  & Application Layer & \impact{3} & \impact{3} & \targeted  & \impact{4}  & \BlackCircle{} & Standard & remote & \mexpert & \MediumFeasibility{} & \ApplyGradient{4} \\
\hline
    \multicolumn{2}{l}{Accuracy:} & \multicolumn{7}{l}{\targeted Targeted, \ntargeted Untargeted}\\
    \multicolumn{2}{l}{Impact:} & \multicolumn{7}{l}{\impact{1} Negligible, \impact{2} Moderate, \impact{3} Major, \impact{4} Severe}\\
    \multicolumn{2}{l}{Knowledge:} & \multicolumn{7}{l}{\WhiteCircle{} white-box, \GrayCircle{} gray-box, \BlackCircle{} black-box}\\
    \multicolumn{2}{l}{Feasibility:} & \multicolumn{7}{l}{\VeryLowFeasibility{} Very Low, \LowFeasibility{} Low, \MediumFeasibility{}  Medium, \HighFeasibility{} High}\\
    \multicolumn{2}{l}{Expertise:} & \multicolumn{7}{l}{\Layman{} Layman, \prof{} Proficient, \expert{} Expert, \mexpert{} Multiple Experts}\\
    \multicolumn{2}{l}{Risk:} & \multicolumn{7}{l}{According to ISO 21434~\cite{international_organization_for_standardization_isosae_2021} from low~\ApplyGradient{1} to high~\ApplyGradient{5}}\\
  \end{tabular}
  \end{adjustbox}
\end{table*}

Current security research on camera pipelines often lacks a consistent risk evaluation, resulting in incomparable or even misleading threat assessments. To address this gap, we systematically evaluate the risk of 48 attack papers (with 8 out of the 56 surveyed papers focusing solely on defense) mapped to the different layers of the pipeline. We assess their impact and feasibility using ISO 21434~\cite{international_organization_for_standardization_isosae_2021}, a well-defined, internationally recognized industry standard specifically designed to address cybersecurity risks in the automotive industry, ensuring a comprehensive and industry-relevant analysis. Details on the threat analysis and risk assessment with ISO 21434 are given in Appendix~\ref{app:riskAnalysisCameraAttacks}. \autoref{tab:comparisonSecurity} presents the attack impact and feasibility rating as well as the final risk level from low~\ApplyGradient{1} to high~\ApplyGradient{5}. To assess the risk, we include the following categories as defined in ISO~21434~\cite{international_organization_for_standardization_isosae_2021}:
\begin{itemize}[nosep]
    \item \textit{Asset Identification}: Characterize the layer of the image processing pipeline where the attack unfolds its potential.
    \item \textit{Attack Path Analysis}: Define the attack entry point used by the attacker.
    \item \textit{Impact Rating}: Rate the impact (\impact{1} Negligible, \impact{2} Moderate, \impact{3} Major, or \impact{4} Severe) of the presented attack according to the following criteria
    \begin{enumerate*}[before=\unskip{: }, itemjoin={{; }}, itemjoin*={{, and }}, label={(\roman*)}]
        \item Impact on safety
        \item Impact on operation
        \item Targeted accuracy, describing whether the result of the attack can be actively controlled (Targeted \targeted{}) or not \mbox{(Untargeted \ntargeted{}).}
    \end{enumerate*}
    \item \textit{Attack Feasibility}: Evaluate the feasibility of the attack (\VeryLowFeasibility{} Very Low, \LowFeasibility{} Low, \MediumFeasibility{} Medium, or \HighFeasibility{} High) based on the following criteria
    \begin{enumerate*}[before=\unskip{: }, itemjoin={{; }}, itemjoin*={{, and }}, label={(\roman*)}]
        \item Required knowledge of the system to perform the attack (\WhiteCircle{} white-box, \GrayCircle{} gray-box, or \BlackCircle{} black-box)
        \item Necessary equipment (Standard, Specialized, Bespoke, or Multiple Bespoke)
        \item Window of opportunity, which refers to the distance between the attacker's system to the victim
        \item Specialist expertise (\Layman{}Layman, \prof{} Proficient, \expert{}Expert, or \mexpert{} Multiple Experts) 
    \end{enumerate*}
\end{itemize}

All ratings in \autoref{tab:comparisonSecurity} are generated with our interactive tool TARA-CAM\footnote{\url{https://tum-esi.github.io/PICT/}}. For each attack, we provide the exact ISO 21434 parameters to enable full reproducibility. Readers can add new attacks or modify any parameter and instantly see the recalculated risk level, enabling exploration of how changes -- such as shifting the window of opportunity -- affect feasibility.

\subsection{Physical World}
All physical-world attacks share a common criterion: their entry point is rooted in the physical environment. However, the risk levels of these attacks can vary widely, from low (\ApplyGradient{1}) to high (\ApplyGradient{5}). Although most attacks have a severe (\impact{4}) or major (\impact{3}) impact, their feasibility goes from very low (\VeryLowFeasibility{})~\cite{patel_overriding_2022, athalye_synthesizing_2018, jing_too_2021, he_dorpatch_2024} to high (\HighFeasibility{})~\cite{lovisotto_slap_2021, nassi_phantom_2020, jia_fooling_2022}. Especially, the required knowledge and expertise contribute to the feasibility of physical world attacks, resulting in varying risks. Except for two papers~\cite{patel_overriding_2022, mohajeransari_discovering_2024}, all attacks in this layer need only standard or specialized equipment. The number of attacks resulting in high risk (\ApplyGradient{4} and \ApplyGradient{5})~\cite{wang_dual_2021, jia_fooling_2022, petit_remote_2015, duan_adversarial_2021, mohajeransari_discovering_2024, gnanasambandam_optical_2021, lovisotto_slap_2021, bhupathiraju_vulnerability_2024, nassi_phantom_2020}, their limited black-box knowledge (\BlackCircle{}) and the possibility of them being targeted (\targeted{}) should guide future defensive research of the image processing pipeline.

\subsection{Sensor Layer}
Similar to the previous layer, nearly all attacks on the sensor layer have their entry point in the physical world, except for~\cite{oyama_adversarial_2024}, which targets the interconnection between the sensor layer and the data preparation layer. Additionally, the risk levels are also distributed from low (\ApplyGradient{1}) to high (\ApplyGradient{5}). In contrast to the physical world, attacks with a very low feasibility (\VeryLowFeasibility{})~\cite{ji_poltergeist_2021, long_side_2023, oyama_adversarial_2024} require different knowledge but unite in a small window of opportunity (<~0.1m). In general, more attacks require bespoke or even multiple bespoke~\cite{oyama_adversarial_2024} equipment compared to attacks in the physical world. The variation in the required knowledge and expertise results in very low (\VeryLowFeasibility{}) to high (\HighFeasibility{}) feasibility. A total of six attacks show only a moderate impact (\impact{2}) but still range up to risk level \ApplyGradient{3}.

\subsection{Data Preparation Layer}
While attacks on preceding layers had their attack entry point mainly in the physical world, here we identified the first attacks with an entry point on the application layer~\cite{li_adversarial_2022, xiao_seeing_2019}. Due to their similar operation principle, all attacks in this layer share a major impact (\impact{3}) and medium feasibility (\MediumFeasibility{}). We emphasize that no work requires specific knowledge, resulting in black-box attacks (\BlackCircle{}). These properties lead to a common risk level of \ApplyGradient{3}.

\subsection{Application Layer}
Many attacks targeting the application layer~\cite{xiong_multi-source_2021, ma_slowtrack_2024, agarwal_crafting_2022} also use it as their entry point, setting them apart from most attacks in other layers. A key limitation of these attacks is that the manipulated images must be accessible, often requiring stepping-stone attacks through other components, which adds complexity. However, exceptions exist, such as those described in~\cite{sato_robustness_2021} and~\cite{boloor_attacking_2020}, which target the application layer while using the physical world as entry points. Irrespective of the entry point, the impact of all attacks ranges from moderate (\impact{2})~\cite{muller_investigating_2025} to severe (\impact{4})~\cite{xiong_multi-source_2021, boloor_attacking_2020, agarwal_crafting_2022}. With a feasibility only ranging from very low (\VeryLowFeasibility{}) to medium (\MediumFeasibility{}), the highest risk level of these attacks is \ApplyGradient{4}.

\observation{Attacks on the physical world and sensor layers often combine high feasibility, little system knowledge, and high impact, often resulting in high risk scores. In contrast, attacks in the data preparation and application layers require high expertise and are harder to execute due to possible prior stepping-stone attacks. Yet, some still score high risk due to their high impact. This cross-layer pattern, where knowledge requirements shift from low to high as one moves up the pipeline, emerges clearly only when threats are mapped systematically.}
\section{Robustness Classification}
\label{sec:robustnessConsiderations}
The development of camera-based autonomous driving solutions has demonstrated a mature level of safe and robust perception, utilizing \ac{CV} algorithms even in high-speed scenarios for nearly four decades~\cite{dickmanns_autonomous_1987}. This progress motivates us to analyze research on the image processing pipeline not only from a security perspective but also by exploring how security can be enhanced through insights from the robustness domain. To achieve this, we reviewed 36 robustness-related studies, with the methodology for selecting and excluding these papers detailed in Appendix~\ref{app:researchmethod}. We map these papers across different layers of the image processing pipeline using multiple criteria and introduce \textbf{six robustness classes} that can be used to categorize the collected research papers. By doing so, we aim to bridge the gap between work in the robustness domain and security-related studies, as discussed in \secref{sec:linkroubandsec}. The result of the classification of 36 papers is presented in \autoref{tab:RobustnessLayer}. The criteria used for classifying the papers include:

\begin{itemize}
    \item \textit{Layer}: We use the layers of the image processing pipeline in \autoref{fig:teaser} to organize the individual subsections. As many papers operate on multiple layers, we also have subsections on these combined ones.
    \item \textit{Contribution}: While some work analyzes and optimizes the existing image processing pipeline components, others aim to identify new structures or working principles. We, therefore, differentiate between \textit{Analysis} (\resizebox{!}{.7em}{\analysis}), analyzing the existing system, \textit{Optimization} (\resizebox{!}{.7em}{\optimization}), aiming to optimize the existing system, and \textit{Upgrades} (\resizebox{!}{.7em}{\updatei}), describing work that aims to use new approaches to increase the robustness.
    \item \textit{Impact}: Since the human visual perception primarily drove the development of camera systems, we differentiate work that optimizes the \textit{Quality} of images or work that focuses on \ac{CV} applications, aiming to improve their \textit{Safety}. There is also research considering both categories.
    \item \textit{Effort}: This represents the effort that is necessary to implement the proposed work in a system. It can be
    \begin{enumerate*}[before=\unskip{: }, itemjoin={{; }}, itemjoin*={{, or }}, label={(\roman*)}]
        \item a \textit{minor software} update (\,\progressbar[0.3]{0.25})
        \item a \textit{major software} update (\,\progressbar[0.3]{0.5})
        \item a necessary \textit{hardware} upgrade (\,\progressbar[0.3]{0.55})
        \item changes that even concern the surrounding \textit{infrastructure} (\,\progressbar[0.3]{1}).
    \end{enumerate*}
\end{itemize}

\begin{table*}[t]
       \setlength{\tabcolsep}{0.5pt}
    \renewcommand{\arraystretch}{1.7}
    \centering
    \caption{Layer classification of robustness-related work: \triangledtext{1} Environmental and data processing analysis, \triangledtext{2} Anti-Flicker improvement, \triangledtext{3} CFA improvement, \triangledtext{4} Low layer sensor improvement, \triangledtext{5} Digital image processing improvement, \triangledtext{6}~Application improvement; \resizebox{!}{.7em}{\analysis} Analysis, \resizebox{!}{.7em}{\optimization} Optimization, \resizebox{!}{.7em}{\updatei} Upgrade}
    \label{tab:RobustnessLayer}
    \setlength{\tabcolsep}{1pt}
    \resizebox{0.95\linewidth}{!}{%
    \begin{tabular}{|*{1}{c}|*{6}{c}|*{11}{c}|*{16}{c}|*{1}{c}|*{2}{c}|}\hline
    \multicolumn{1}{|c|}{\textbf{Layer}} & \multicolumn{6}{c|}{\textbf{Physical/Sensor}} & \multicolumn{11}{c|}{\textbf{Sensor}} & \multicolumn{16}{c|}{\textbf{Sensor/Data Prep.}} & \multicolumn{1}{c|}{\textbf{Data}} & \multicolumn{2}{c|}{\textbf{App.}} \\\hline
    \multicolumn{1}{|c|}{\textbf{Work}} & \rotatebox[origin=c]{90}{\cite{ceccarelli_rgb_2023}} & \rotatebox[origin=c]{90}{\cite{bijelic_benchmarking_2018}} & \rotatebox[origin=c]{90}{\cite{hendrycks_benchmarking_2019}} & \rotatebox[origin=c]{90}{\cite{temel_cure-or_2018}} & \rotatebox[origin=c]{90}{\cite{haralick_performance_1992}} & \rotatebox[origin=c]{90}{\cite{ji_error_1999}} & \rotatebox[origin=c]{90}{\cite{bielova_digital_2019}} & \rotatebox[origin=c]{90}{\cite{deegan_led_2018}} & \rotatebox[origin=c]{90}{\cite{behmann_selective_2018}} & \rotatebox[origin=c]{90}{\cite{silsby_12mp_2015}} & \rotatebox[origin=c]{90}{\cite{funatsu_non-rgb_2022}} & \rotatebox[origin=c]{90}{\cite{weikl_optimization_2020}} & \rotatebox[origin=c]{90}{\cite{chakrabarti_learning_2016}} & \rotatebox[origin=c]{90}{\cite{cote_differentiable_2023}} & \rotatebox[origin=c]{90}{\cite{mann_being_1994}} & \rotatebox[origin=c]{90}{\cite{vera_shuffled_2022}} & \rotatebox[origin=c]{90}{\cite{sonoda_high-speed_2016}} & \rotatebox[origin=c]{90}{\cite{dodge_understanding_2016}} & \rotatebox[origin=c]{90}{\cite{karahan_how_2016}} & \rotatebox[origin=c]{90}{\cite{yahiaoui_impact_2018}} & \rotatebox[origin=c]{90}{\cite{pezzementi_putting_2018}} & \rotatebox[origin=c]{90}{\cite{blasinski_optimizing_2018}} & \rotatebox[origin=c]{90}{\cite{yahiaoui_overview_2019}} & \rotatebox[origin=c]{90}{\cite{mosleh_hardware---loop_2020}} & \rotatebox[origin=c]{90}{\cite{molloy_impact_2023}} & \rotatebox[origin=c]{90}{\cite{yahiaoui_optimization_2019}} & \rotatebox[origin=c]{90}{\cite{robidoux_end--end_2021}} & \rotatebox[origin=c]{90}{\cite{wang_adaptiveisp_2024}} & \rotatebox[origin=c]{90}{\cite{heide_flexisp_2014}} & \rotatebox[origin=c]{90}{\cite{buckler_reconfiguring_2017}} & \rotatebox[origin=c]{90}{\cite{karam_optimizing_2024}} & \rotatebox[origin=c]{90}{\cite{diamond_dirty_2021}} & \rotatebox[origin=c]{90}{\cite{yu_reconfigisp_2021}} & \rotatebox[origin=c]{90}{\cite{aqqa_understanding_2019}} & \rotatebox[origin=c]{90}{\cite{zhou_classification_2017}} & \rotatebox[origin=c]{90}{\cite{lou_dc-yolov8_2023}}\\ \hline
    \multicolumn{1}{|c|}{\textbf{Class}} & \triangledtext[-0.5ex]{1} & \triangledtext[-0.5ex]{1} & \triangledtext[-0.5ex]{1} & \triangledtext[-0.5ex]{1} & \triangledtext[-0.5ex]{1} & \triangledtext[-0.5ex]{1} & \triangledtext[-0.5ex]{1} & \triangledtext[-0.5ex]{2} & \triangledtext[-0.5ex]{2} & \triangledtext[-0.5ex]{2} & \triangledtext[-0.5ex]{3} & \triangledtext[-0.5ex]{3} & \triangledtext[-0.5ex]{3} & \triangledtext[-0.5ex]{4} & \triangledtext[-0.5ex]{4} & \triangledtext[-0.5ex]{4} & \triangledtext[-0.5ex]{4} & \triangledtext[-0.5ex]{1} & \triangledtext[-0.5ex]{1} & \triangledtext[-0.5ex]{5} & \triangledtext[-0.5ex]{5} & \triangledtext[-0.5ex]{5} & \triangledtext[-0.5ex]{5} & \triangledtext[-0.5ex]{5} & \triangledtext[-0.5ex]{5} & \triangledtext[-0.5ex]{5} & \triangledtext[-0.5ex]{5} & \triangledtext[-0.5ex]{5} & \triangledtext[-0.5ex]{5} & \triangledtext[-0.5ex]{5} & \triangledtext[-0.5ex]{5} & \triangledtext[-0.5ex]{5} & \triangledtext[-0.5ex]{5} & \triangledtext[-0.5ex]{1} & \triangledtext[-0.5ex]{6} & \triangledtext[-0.5ex]{6} \\ \hline
    \multicolumn{1}{|c|}{\textbf{Contr.}} & \analysis & \analysis & \analysis & \analysis & \analysis & \analysis & \optimization & \analysis & \optimization & \updatei & \optimization & \optimization & \updatei & \updatei & \updatei & \updatei & \updatei & \analysis & \analysis & \analysis & \analysis & \optimization & \optimization & \optimization & \optimization & \optimization & \optimization & \updatei & \updatei & \updatei & \updatei & \updatei & \updatei & \analysis & \updatei & \updatei \\ \hline  
    \multicolumn{1}{|c|}{\textbf{Safety}} & \cmark & \cmark & \cmark & \cmark &  &  & \cmark &  & \cmark & \cmark &  &  & \cmark & \cmark &  &  & \cmark & \cmark & \cmark & \cmark & \cmark &  & \cmark & \cmark & \cmark & \cmark & \cmark & \cmark &  & \cmark & \cmark & \cmark & \cmark & \cmark & \cmark & \cmark \\
    \multicolumn{1}{|c|}{\textbf{Quality}}   &   &   &   & \cmark & \cmark &   & \cmark & \cmark & \cmark & \cmark & \cmark & \cmark &   & \cmark & \cmark & \cmark &   &   &   &   & \cmark &   &   &   &   &   &   & \cmark &   &   & \cmark & \cmark &   &   &  & \\ \hline
    \multicolumn{1}{|c|}{\textbf{Effort}} & \progressbar[0.4]{0.75} & \progressbar[0.4]{0.75} & \progressbar[0.4]{0.5} & \progressbar[0.4]{0.5} & \progressbar[0.4]{0.25} & \progressbar[0.4]{0.25} & \progressbar[0.4]{0.5} & \progressbar[0.4]{1} & \progressbar[0.4]{1} & \progressbar[0.4]{0.75} & \progressbar[0.4]{0.75} & \progressbar[0.4]{0.75} & \progressbar[0.4]{1} & \progressbar[0.4]{0.75} & \progressbar[0.4]{1} & \progressbar[0.4]{1} & \progressbar[0.4]{1} & \progressbar[0.4]{0.25} & \progressbar[0.4]{0.25} & \progressbar[0.4]{0.25} & \progressbar[0.4]{0.25} & \progressbar[0.4]{0.5} & \progressbar[0.4]{0.25} & \progressbar[0.4]{0.5} & \progressbar[0.4]{0.25} & \progressbar[0.4]{0.25} & \progressbar[0.4]{0.5} & \progressbar[0.4]{0.5} & \progressbar[0.4]{1} & \progressbar[0.4]{1} & \progressbar[0.4]{1} & \progressbar[0.4]{1} & \progressbar[0.4]{0.75} & \progressbar[0.4]{0.5} & \progressbar[0.4]{0.5} & \progressbar[0.4]{0.5} \\ \hline
    \end{tabular}
    }
\end{table*}

\subsection{Physical World / Sensor Layer}

\cpar{\triangledtext{1}~Environmental and Data Processing Analysis}
In general, work that acts both in the physical world and in the sensor layer can be allocated to this class, where some work analyzes the effect of environmental influences and sensor distortions~\cite{ceccarelli_rgb_2023, hendrycks_benchmarking_2019}, or even provide datasets for evaluation~\cite{temel_cure-or_2018}, and other work explains on a more theoretical level the performance of \ac{CV} algorithms under the influence of random changes~\cite{haralick_performance_1992, ji_error_1999}. In a different analysis, \citeauthor{bijelic_benchmarking_2018}~\cite{bijelic_benchmarking_2018} investigate environmental effects on both standard \ac{CMOS} cameras and so-called gated cameras. This supporting work on more robust \ac{CV} algorithms can also increase the robustness against adversarial distortions both in the real world and in the application layer.

\subsection{Sensor Layer}

In contrast to the previous subsection, here we focus exclusively on research related to the sensor layer. We identified three defense classes: \triangledtext{2}~\textit{Anti-Flicker Improvements}, \triangledtext{3}~\textit{\ac{CFA} Improvements}, and \triangledtext{4}~\textit{Low-Layer Sensor Improvements}.

\cpar{\triangledtext{2}~Anti-Flicker Improvements}
Various methods exist to detect and mitigate the flickering effect of pulsed \acp{LED}~\cite{deegan_led_2018, behmann_selective_2018, silsby_12mp_2015}. 
Many automotive image sensors already support \ac{LED} flicker mitigation, often using proprietary algorithms~\cite{on_semiconductor_hayabusa_2024}. Since these algorithms are often integrated directly on the chip, they typically require a high implementation effort, although it is also possible to implement such functions as part of the data preparation in a separate component~\cite{behmann_real-time_2018}.

\cpar{\triangledtext{3}~\ac{CFA} Improvements}
This class focuses on improvements of the \ac{CFA} for automotive-specific requirements, which can influence the outcome of attacks that use digital images or targeted light. Research has examined \acp{CFA} tailored to the automotive industry~\cite{weikl_optimization_2020, chakrabarti_learning_2016, funatsu_non-rgb_2022}. While~\cite{weikl_optimization_2020} and~\cite{funatsu_non-rgb_2022} explore the best existing \ac{CFA} for traffic signal detection, \citeauthor{chakrabarti_learning_2016}~\cite{chakrabarti_learning_2016} proposes replacing the traditional Bayer \ac{CFA} with an adaptive learned \ac{CFA}.

\cpar{\triangledtext{4}~Low-Layer Sensor Improvements}
A multitude of improvements of the image sensor is available. These improvements can include advanced lens stacks, which are crucial for creating high-quality images and can significantly boost the performance of object detection algorithms~\cite{cote_differentiable_2023}.
Additionally, \ac{HDR} image sensors can play a vital role in defending against simple blinding attacks through targeted light by sequentially capturing multiple images with varying exposure settings to create a single image with a higher dynamic range~\cite{mann_being_1994}. 
Although potentially introducing motion artifacts on fast-moving objects, image sensors with such an \ac{HDR} feature onboard are common in the automotive industry~\cite{samsung_semiconductor_isocell_2024}. Another area of low-layer sensor improvements aims to reduce the rolling shutter effect, which is particularly important in high-speed scenarios~\cite{vera_shuffled_2022, sonoda_high-speed_2016}. Furthermore, sensor hardware properties can be leveraged to generate more realistic noise in artificial images~\cite{bielova_digital_2019}, which can be utilized in technologies such as training data randomization~\cite{xie_mitigating_2018} or adversarial training~\cite{madry_towards_2018}. While much of the research is focused on optimizing image quality for the human visual system, some studies also enhance the performance of \ac{CV} algorithms~\cite{behmann_selective_2018, silsby_12mp_2015, cote_differentiable_2023, bielova_digital_2019}. Since optimizations and upgrades to the sensor layer often involve hardware changes or even infrastructural modifications, the effort for work in this layer is typically high.

\observation{Although robustness-related improvements in the sensor layer can enhance the security of the image processing pipeline, they often come with a high effort, affecting not only the hardware itself but also its infrastructure. This makes a thorough evaluation from a security perspective difficult and shows the need for a camera testbed featuring realistic automotive features.}

\subsection{Sensor Layer / Data Preparation Layer}
Two classes were identified: \triangledtext{1}~\textit{Environmental and Data Processing Analysis}, and \triangledtext{5}~\textit{Digital Image Processing Improvements}.

\cpar{\triangledtext{1}~Environmental and Data Processing Analysis}
Different research focused on analyzing various distortions within the image processing and their impact on neural networks~\cite{dodge_understanding_2016, karahan_how_2016} and perception algorithms~\cite{pezzementi_putting_2018}. The analyzed distortions relate to digital images, printed objects, or adversarial distortions from security-related research.

\cpar{\triangledtext{5}~Digital Image Processing Improvements}
Improvements in this class aim to ameliorate digitized image data processing. Since the exact functionalities of the on-chip processing in the sensor layer and the functions of the \ac{ISP} within the data preparation layer are not standardized, we will classify the respective work in this subsection, covering both layers. Much of this research aims to optimize parameters of the \ac{ISP}~\cite{blasinski_optimizing_2018, yahiaoui_overview_2019, robidoux_end--end_2021, mosleh_hardware---loop_2020, yahiaoui_optimization_2019, yahiaoui_impact_2018, pezzementi_putting_2018, molloy_impact_2023}, but due to the possible tasks of their proposed work, it is classified in this combined subsection. Similar work~\cite{heide_flexisp_2014, buckler_reconfiguring_2017, diamond_dirty_2021, yu_reconfigisp_2021, wang_adaptiveisp_2024} does not only try to optimize parameters on existing components both in the sensor layer and data preparation layer but tries to replace components of the image processing pipeline with new ones to improve the visual quality of images and the performance of \ac{CV} algorithms. By optimizing or replacing processing steps in the image processing pipeline, the impact of adversarial distortions, printed objects, or other physical attacks can potentially be reduced. A similar approach is given by \citeauthor{avidan_all_2022}~\cite{avidan_all_2022}. 

By considering light outside the human visible spectrum, both the image sensor and data processing algorithms can be optimized to improve the performance of pedestrian detection algorithms~\cite{karam_optimizing_2024}. Nevertheless, such major changes typically require more effort than software-based image processing pipeline modifications.

\subsection{Data Preparation Layer}

\cpar{\triangledtext{1}~Environmental and Data Processing Analysis}
Nearly all basic image processing functions can be integrated with on-chip processing in the sensor layer, as the processing steps between these layers often overlap. Only more complex image transformations, such as image compression algorithms, are typically handled by the more capable \ac{ISP}. Therefore, analyzing image compression algorithms and their impact on object detection algorithms~\cite{aqqa_understanding_2019} is crucial in defending against attacks on image transformations~\cite{agarwal_crafting_2022}.

\subsection{Application Layer}
\label{sec:robustnessApplicationLayer}

\cpar{\triangledtext{6}~Application Improvements}
All research in this layer aims to enhance the software algorithms. Similar to the security-related research on the application layer discussed in~\secref{sec:securityWorkApplication}, we explicitly exclude work that does not involve components of the image processing pipeline or is not specifically applicable to autonomous driving. Existing surveys and reviews~\cite{liu_deep_2020, zhao_object_2019, zhang_towards_2020} provide an overview of methods to create robust \ac{CV} algorithms that operate on the application layer.

Enhancing the performance of image detectors~\cite{lou_dc-yolov8_2023} can improve the robustness of object detection algorithms in autonomous vehicles, particularly in scenarios where vehicles are far away. Another approach to increase robustness on the application layer is improving classification algorithms through fine-tuning with noisy data~\cite{zhou_classification_2017}, which could be further enhanced with more realistic image sensor noise simulations~\cite{bielova_digital_2019}. Both methods share the potential to counteract adversarial perturbations, whether through digital images, printed objects, or adversarial distortions in the digital domain.
\section{Linking Robustness and Security Work}
\label{sec:linkroubandsec}

\newcolumntype{?}{!{\color{gray!30}\vrule width 1pt}}
\newcolumntype{Q}{!{\vrule width 2pt}}

\begin{table*}[t]
\centering
\caption{Comparing security- and robustness-related research work. Many works act on similar operational principles. As highlighted by the green references, some attack classes received additional research from defensive security work.}
\label{tab:linkSecurityRobustness}

\resizebox{0.95\textwidth}{!}{%
\begin{tabular}{|cccQc?c?c|c?c?c?c?c|c|c?c?c|}
\hline
\multicolumn{3}{|cQ}{\textbf{Layer}} & \multicolumn{3}{c|}{\textbf{Physical}} & \multicolumn{5}{|c|}{\textbf{Sensor}} & \multicolumn{1}{|c|}{\textbf{Data Prep.}} & \multicolumn{3}{|c|}{\textbf{App.}} \\ \cline{2-15}
 & \multicolumn{1}{|c|}{\textbf{Class}} & \multicolumn{1}{|cQ}{\textbf{Security}} & \circledtexted{1} & \circledtexted{2} & \circledtexted{3} & \circledtexted{1} & \circledtexted{3} & \circledtexted{4} & \circledtexted{5} & \circledtexted{6} & \circledtexted{5} & \circledtexted{5} & \circledtexted{7} & \cellcolor{red!10}\circledtexted{8} \\ \cline{2-15}
 %
 %
 & \multicolumn{1}{|c|}{\textbf{Robustness}} & \multicolumn{1}{|cQ}{\textbf{Work}} & 
 \begin{tabular}[c]{@{}l@{}}
     \cite{patel_overriding_2022} \cite{gnanasambandam_optical_2021} \\ \cite{nassi_phantom_2020} \cite{lovisotto_slap_2021} 
 \end{tabular} & \begin{tabular}[c]{@{}l@{}}
     \cite{athalye_synthesizing_2018} \cite{eykholt_robust_2018} \cite{radutoiu_physical_2023} \cite{jan_connecting_2019} \\ \cite{kong_physgan_2020} \cite{wang_dual_2021} \cite{huang_universal_2020} \cite{he_dorpatch_2024} \\ \cite{jing_too_2021} \cite{jia_fooling_2022} \cite{zhao_seeing_2019} \cite{mohajeransari_discovering_2024} 
 \end{tabular} & 
 \begin{tabular}[c]{@{}l@{}}
     \cite{petit_remote_2015} \cite{bhupathiraju_vulnerability_2024} \\ \cite{duan_adversarial_2021} \cite{hsiao_natural_2024} \\
     \cite{wang_rfla_2023}
 \end{tabular} & 
 \begin{tabular}[c]{@{}l@{}}
     \cite{xia_moire_2024} \\ \cite{man_remote_2024} 
 \end{tabular}
 &
 \begin{tabular}[c]{@{}l@{}}
     \cite{zhou_doublestar_2022} \\ \cite{sato_invisible_2024} \\ \cite{wang_i_2021} 
 \end{tabular}
 & 
 \begin{tabular}[c]{@{}l@{}}
     \cite{ji_poltergeist_2021} \\ \cite{kohler_signal_2022} \\ \cite{long_side_2023} 
 \end{tabular}
 &
 \begin{tabular}[c]{@{}l@{}}
     \cite{oyama_adversarial_2024} 
 \end{tabular}
 & 
 \begin{tabular}[c]{@{}l@{}}
     \cite{huang_lights_2022} \cite{wu_illumination_2021} \cite{sayles_invisible_2021} \cite{guo_invisible_2024} \\ \cite{li_light_2020} \cite{yan_rolling_2022} \cite{kohler_they_2021} \cite{liu_cross-task_2024}
 \end{tabular}
 & 
 \begin{tabular}[c]{@{}l@{}}
     \cite{phan_adversarial_2021} \cite{li_adversarial_2022} \\ \cite{li_image-scaling_2024} \cite{xiao_seeing_2019} 
 \end{tabular}
 & 
 \begin{tabular}[c]{@{}l@{}}
     \cite{agarwal_crafting_2022} 
 \end{tabular}
 & 
 \begin{tabular}[c]{@{}l@{}}
     \cite{xiong_multi-source_2021} \\ \cite{boloor_attacking_2020} \\ \cite{sato_robustness_2021} 
 \end{tabular}
 &
 \cellcolor{red!10}\begin{tabular}[c]{@{}l@{}}
     \cite{ma_slowtrack_2024} \\ \cite{muller_investigating_2025} 
 \end{tabular} \\ \specialrule{2.0pt}{0pt}{0pt}
\raisebox{2mm}{\multirow{1}{*}{\begin{tabular}[c]{@{}c@{}}\textbf{Physical/}\\\textbf{Sensor}\end{tabular}}} & \multicolumn{1}{|c|}{\triangledtext{1}} & \begin{tabular}[c]{@{}l@{}}
     \cite{ceccarelli_rgb_2023} \cite{bijelic_benchmarking_2018} \\ \cite{hendrycks_benchmarking_2019} \cite{temel_cure-or_2018} \\ \cite{haralick_performance_1992} \cite{ji_error_1999} 
 \end{tabular} & \begin{tabular}[c]{@{}l@{}}
     \OK \\ \textcolor{black!60!green}{\cite{feng_fight_2025}}
 \end{tabular} & \begin{tabular}[c]{@{}l@{}}
     \OK \\ \textcolor{black!60!green}{\cite{feng_fight_2025}}
 \end{tabular} & \OK & & & \OK & & & & & \OK & \cellcolor{red!10} \\ \specialrule{1.0pt}{0pt}{0pt}
\multirow{9}{*}{\rotatebox[origin=c]{90}{\begin{tabular}[c]{@{}c@{}}\textbf{Sensor}\end{tabular}}} & \multicolumn{1}{|c|}{\triangledtext{1}} & \begin{tabular}[c]{@{}l@{}}
     \cite{bielova_digital_2019}
 \end{tabular} & & & & \OK & & & & & & & & \cellcolor{red!10} \\ \Cline{2-15}
 & \multicolumn{1}{|c|}{\triangledtext{2}} & \begin{tabular}[c]{@{}l@{}}
     \cite{deegan_led_2018} \\ \cite{behmann_selective_2018} \\ \cite{silsby_12mp_2015}
 \end{tabular} & & & & & & & & \OK & & & & \cellcolor{red!10} \\ \Cline{2-15}
 & \multicolumn{1}{|c|}{\triangledtext{3}} & \begin{tabular}[c]{@{}l@{}}
     \cite{funatsu_non-rgb_2022} \\ \cite{weikl_optimization_2020} \\ \cite{chakrabarti_learning_2016}
 \end{tabular} & \OK & & & & \OK & & & & & & & \cellcolor{red!10} \\ \Cline{2-15}
 & \multicolumn{1}{|c|}{\triangledtext{4}} & \begin{tabular}[c]{@{}l@{}}
     \cite{cote_differentiable_2023} \cite{mann_being_1994} \\ \cite{vera_shuffled_2022} \cite{sonoda_high-speed_2016}
 \end{tabular} & \OK & & \OK & & & & & \OK & & & & \cellcolor{red!10} \\ \specialrule{1.0pt}{0pt}{0pt}
\multirow{6}{*}{\rotatebox[origin=c]{90}{\begin{tabular}[c]{@{}c@{}}\textbf{Sensor./}\\\textbf{Data Prep.}\end{tabular}}} & \multicolumn{1}{|c|}{\triangledtext{1}} & \begin{tabular}[c]{@{}l@{}}
     \cite{karahan_how_2016} \\ \cite{dodge_understanding_2016}
 \end{tabular} & \OK & \OK & \OK & & & & & & & & \OK & \cellcolor{red!10} \\ \Cline{2-15}
 & \multicolumn{1}{|c|}{\triangledtext{5}} & \begin{tabular}[c]{@{}l@{}}
     \cite{yahiaoui_impact_2018} \cite{pezzementi_putting_2018} \cite{blasinski_optimizing_2018} \\
     \cite{yahiaoui_overview_2019} \cite{mosleh_hardware---loop_2020} \cite{molloy_impact_2023} \\
     \cite{robidoux_end--end_2021} \cite{wang_adaptiveisp_2024} \cite{heide_flexisp_2014}\\
     \cite{karam_optimizing_2024} \cite{diamond_dirty_2021} \cite{yu_reconfigisp_2021}\\
     \cite{yahiaoui_optimization_2019} \cite{buckler_reconfiguring_2017}
 \end{tabular} & \OK & \OK & \OK & & & \OK & \textcolor{black!60!green}{\cite{mipi_alliance_guide_2024}} & & \begin{tabular}[c]{@{}l@{}}
     \OK \\ \textcolor{black!60!green}{\cite{quiring_adversarial_2020}} \\ \,\textcolor{black!60!green}{\cite{mipi_alliance_guide_2024}}
 \end{tabular} & \textcolor{black!60!green}{\cite{mipi_alliance_guide_2024}} & \textcolor{black!60!green}{\cite{mipi_alliance_guide_2024}} & \cellcolor{red!10} \\ \specialrule{1.0pt}{0pt}{0pt}
\multirow{1}{*}{\begin{tabular}[c]{@{}c@{}}\textbf{Data Prep.}\end{tabular}} & \multicolumn{1}{|c|}{\triangledtext{1}} & \begin{tabular}[c]{@{}l@{}}
     \cite{aqqa_understanding_2019}
 \end{tabular} & \textcolor{black!60!green}{\cite{avidan_all_2022}} & \textcolor{black!60!green}{\cite{avidan_all_2022}} & \textcolor{black!60!green}{\cite{avidan_all_2022}} & & & & & & & \OK & \textcolor{black!60!green}{\cite{avidan_all_2022}} & \cellcolor{red!10} \\ \specialrule{1.0pt}{0pt}{0pt}
\multirow{1}{*}{\begin{tabular}[c]{@{}c@{}}\textbf{App.}\end{tabular}} & \multicolumn{1}{|c|}{\triangledtext{6}} & \begin{tabular}[c]{@{}l@{}}
     \cite{lou_dc-yolov8_2023} \cite{zhou_classification_2017}
 \end{tabular} & \begin{tabular}[c]{@{}l@{}}
     \quad\OK \\ \textcolor{black!60!green}{\cite{gurel_knowledge_2021, luo_foveation-based_2016, man_that_2023}}
 \end{tabular} & \begin{tabular}[c]{@{}l@{}}
     \quad\OK \\ \textcolor{black!60!green}{\cite{gurel_knowledge_2021, luo_foveation-based_2016, man_that_2023}}
 \end{tabular} & \textcolor{black!60!green}{\cite{gurel_knowledge_2021, zhang_detecting_2021}} & & \textcolor{black!60!green}{\cite{zhang_detecting_2021}} & & & & & & \begin{tabular}[c]{@{}l@{}}
     \quad\OK \\ \textcolor{black!60!green}{\cite{gurel_knowledge_2021, luo_foveation-based_2016, man_that_2023}}
 \end{tabular} & \cellcolor{red!10} \\ \specialrule{1.0pt}{0pt}{0pt}

\end{tabular}
}

\afterfigspace
\end{table*}

Security and robustness research on camera pipelines has largely evolved in parallel, with little awareness of each other’s advances. As a result, several well-known attacks have been discussed in security literature without consistently acknowledging that relevant mitigations were already available, sometimes for decades. For example, camera blinding attacks on automotive systems, first reported in 2015~\cite{petit_remote_2015}, were often examined without reference to \ac{HDR} imaging, a technique introduced in 1994~\cite{mann_being_1994} to better capture high dynamic range scenes. Similarly, rolling shutter exploits leveraging flickering light~\cite{sayles_invisible_2021} appeared after the introduction of LED flicker mitigation in automotive image sensors~\cite{deegan_led_2018, behmann_selective_2018, silsby_12mp_2015}, yet this defense is not addressed in security papers. Our \ac{SoK} links robustness and security research, uncovering overlooked connections and enabling both communities to adapt proven solutions. We make these connections visible by mapping attack classes from \autoref{tab:SecurityLayer} to robustness mechanisms from \autoref{tab:RobustnessLayer} in \autoref{tab:linkSecurityRobustness}. This shows not only where existing robustness measures align with security needs, but also where countermeasures remain absent and which layers have the highest overlap. The layers defined in \autoref{fig:teaser} are used to distinguish where the work operates.

Particularly adversarial distortions in the physical world (\circledtexted{1} and \circledtexted{2}) and application layer (\circledtexted{7}) not only received defenses in the security domain but also show high similarities with existing robustness research working on several layers. Since these attack classes rely on precisely calculated distortions, different robustness-related research can be applied to possibly mitigate the impact of such attacks. Similarly, blinding through targeted light (\circledtexted{3}) is not only defended from a security perspective~\cite{zhang_detecting_2021} but also closely related to existing analyses (\triangledtext{1}) and upgrades (\triangledtext{4} and \triangledtext{5}) from a robustness perspective. Additionally, well-known robustness measures, such as light flicker mitigation technologies (\triangledtext{2}), can help mitigate rolling-shutter attacks (\circledtexted{6}) that exploit the sensor sensitivity to fast flickering light. Such close relations show that the whole image processing pipeline must be considered when crafting both attacks and defenses on camera systems. Another observation is the distribution of the work among the layers: The data preparation layer received significantly more robustness-related research, while the security research showed more available work in the physical world. By contrast, highly specialized, recent timing attacks~\cite{muller_investigating_2025, ma_slowtrack_2024} neither have directly related countermeasures nor show strong similarities to existing robustness-related research.

The consideration of robustness-related research and the complete image processing pipeline is necessary, since existing security industry standards~\cite{mipi_alliance_guide_2024} can protect only a fraction of possible attacks. Especially attacks that originate in the physical world (\circledtexted{1}, \circledtexted{2}, and \circledtexted{3}), or attacks that exploit specific sensor characteristics (\circledtexted{4}, and \circledtexted{6}) are still possible with state-of-the-art security measures of industry standards.

\observation{Robustness research on the image processing pipeline is unevenly distributed: many works concentrate on the sensor and data preparation layers, often with high-effort hardware upgrades, while the physical world and application layers see less direct attention. Certain approaches, particularly environmental and processing analyses, appear in multiple layers, reflecting their potential to thwart attacks across the pipeline.}
\section{Image Processing Pipeline Testbed}
\label{sec:imageSensorTestbed}

To leverage the insights from our SoK, we built PICT, an image processing pipeline testbed designed for academic experimentation. PICT is not intended to replicate the full complexity of commercial multi-camera stacks, but to embody the complexity and extensive configuration options of the image processing pipeline identified as missing in existing tools. It allows for manipulating image processing pipeline parameters and investigating intermediate images, contributing to both security research from \autoref{tab:SecurityLayer} and robustness research of camera systems from \autoref{tab:RobustnessLayer}. Moreover, it facilitates developing and testing attacks and countermeasures on the image processing pipeline by enabling controlled image capture with defined settings. Despite the significant demand for such a platform, to our knowledge, no open testbed simultaneously offers realistic hardware, comprehensive pipeline-layer coverage, and reproducibility.

Many existing studies, such as~\cite{agarwal_crafting_2022, athalye_synthesizing_2018, li_image-scaling_2024, ceccarelli_rgb_2023, chakrabarti_learning_2016, chen_understanding_2020, cote_differentiable_2023, dodge_understanding_2016, xiong_multi-source_2021, gurel_knowledge_2021}, have conducted evaluations without fully utilizing hardware testbeds, relying instead on existing datasets or simulations~\cite{blasinski_optimizing_2018, boloor_attacking_2020, ma_slowtrack_2024, patel_overriding_2022, radutoiu_physical_2023}. On the other hand, real-world evaluations involving hardware testbeds often use one of two types of image-capturing devices:
\begin{enumerate*}[before=\unskip{ }, itemjoin={{; }}, itemjoin*={{, or }}, label={(\roman*)}]
    \item \textit{Smartphone cameras}~\cite{cao_invisible_2021, avidan_all_2022, kohler_they_2021, sayles_invisible_2021, long_side_2023, liu_cross-task_2024, yan_rolling_2022, li_light_2020, wu_illumination_2021, jan_connecting_2019, zhao_seeing_2019, ji_poltergeist_2021, eykholt_robust_2018}
    \item \textit{(Semi-) Professional cameras}~\cite{temel_cure-or_2018, molloy_impact_2023, diamond_dirty_2021, yu_reconfigisp_2021, bielova_digital_2019, avidan_all_2022, li_adversarial_2022, gnanasambandam_optical_2021}.
\end{enumerate*} 
Finally, existing testbeds for interacting with image processing pipeline layers do not represent state-of-the-art technology~\cite{rowe_cmucam3_2007, adams_frankencamera_2010}. These observations motivate the development of PICT, a testbed with embedded hardware and software for evaluating multiple pipeline parameters.

\cpar{Design Considerations}
Based on the limitations of existing solutions, we outline key design considerations of PICT:
\begin{enumerate}[nosep]
    \item \textit{Representative image sensor}: It should feature realistic sensor sizes, as these impact image quality. For example, the iPhone 8 Plus as used in~\cite{cao_invisible_2021} has a sensor size of 32.3~mm\textsuperscript{2}~\cite{daniel_yang_apple_2017}, whereas the EOS Rebel T6i, used in~\cite{gnanasambandam_optical_2021}, has a much larger sensor of 332.3 mm\textsuperscript{2}~\cite{digital_camera_database_canon_2024}. As a reference, automotive sensors are typically smaller with a sensor size of $\approx$ 25 mm\textsuperscript{2}~\cite{samsung_semiconductor_isocell_2024, omnivision_technologies_ox03f10_2022, on_semiconductor_hayabusa_2024}. Lastly, it should cover a wide field of view of $\geq$100°, similar to real automotive cameras~\cite{zf_friedrichshafen_ag_smart_2024, continental_automotive_technologies_gmbh_continental_2024, valeo_valeo_2024, robert_bosch_gmbh_multi_2024, mobileye_technologies_limited_enhanced_2022}.
    \item \textit{Use case consideration}: PICT should be able to configure settings that align with automotive \ac{CV} algorithms rather than visually appealing images. Features like advanced lens systems, including optical zoom, which are uncommon or unavailable in automotive systems, should not be considered.
    \item \textit{Modularity and hardware access}: PICT should allow for low-level access to sensor interconnections, which is often restricted in fully assembled devices. This enables the usage of alternative image sensors for specific requirements.
    \item \textit{Consideration of all layers}: PICT must encompass all layers of the image processing pipeline, as depicted in \autoref{fig:teaser}. Current testbeds often overlook lower layers, such as the sensor layer~\cite{leibe_software_2016}.
\end{enumerate}

\cpar{Hardware Architecture}
PICT uses a \textit{Camera Module 3 Wide} with a Sony IMX708 image sensor for evaluation and development with max. 11.9MP and a 102° field of view~\cite{raspberry_pi_ltd_camera_2024}. This camera module supports real-world automotive features, such as an \ac{HDR} mode, and many settings that can be controlled manually. We use a \textit{Raspberry Pi} as an embedded device with the MIPI-\ac{CSI-2} camera interface that is also used in automotive systems on chip~\cite{renesas_electronics_europe_gmbh_renesas_2021}. The Raspberry Pi is solely selected for its ability to provide an easily configurable environment that aligns with the purpose of the testbed, making it an appropriate choice for our investigation. It is not part of the performance characteristics of the image processing pipeline. With a hardware investment of $\approx \$70$, this enables researchers to cost-effectively analyze the image processing pipeline. At the same time, the use of standardized interfaces and easily available hardware components allows modifications, such as adding other sensors.

\cpar{Software Architecture}
To capture and evaluate images, we offer three software tools that enable interaction with different modules of the image processing pipeline. The modular design and standardized data formats facilitate integration with other tools:
\begin{enumerate}[nosep]
    \item \textit{Image Capturing}: Allows the capturing of images both before and after the \ac{ISP}, including the configuration of the on-chip processing and the \ac{ISP} parameters. Since the camera module transmits only one image to the Raspberry Pi, both images can be compared since there is no delay or second exposure.
    \item \textit{Low-Level Analysis}: Enables a manual inspection and analysis of two captured images. It allows researchers to investigate the effect of parameter changes on the captured image. Besides visual inspection, it calculates statistical metrics.
    \item \textit{Application-specific Analysis}: We provide a Faster R-CNN-based object detection tool as an example software~\cite{ren_faster_2015}. By reading both the pre-\ac{ISP} and post-\ac{ISP} image files, PICT allows for the analysis of how individual parameters impact object detection. Additionally, PICT supports custom tools due to the use of standardized data formats.
\end{enumerate}

\section{Evaluation}
\label{sec:pictevaluation}

We will demonstrate the usability and benefits of PICT as well as the influence of the holistic consideration of the image processing pipeline by
\begin{enumerate*}[before=\unskip{: }, itemjoin={{; }}, itemjoin*={{, and }}, label={(\roman*)}]
    \item showcasing PICT's ability to test proposed defense mechanisms of \autoref{tab:linkSecurityRobustness}
    \item providing a better understanding of attacks through analysis capabilities and supporting a more accurate realization of attack scenarios
<    \item highlighting PICT's impact by comparing its result with unrealistic mobile phone cameras, commonly used in research.
\end{enumerate*}
We selected representative attacks of different classes, but other attacks can be evaluated similarly.

\subsection{Defensive Use Case Analysis}
\label{sec:useCaseAnalysis}

We use PICT to evaluate \ac{HDR} imaging~\cite{mann_being_1994, willassen_1280x1080_2015} as a mitigation for camera blinding attacks~\cite{petit_remote_2015}, an example of the overlooked robustness–security connections discussed in \secref{sec:linkroubandsec}. While camera blinding poses a significant risk, our test shows that \ac{HDR} functionality reduces the effect of highly saturated regions in the image.

\begin{figure}[t!]
\centering
    \begin{subfigure}[b]{0.32\linewidth}
         \centering
         \includegraphics[width=0.55\textwidth]{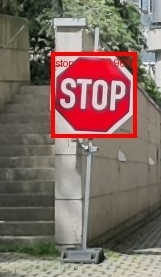}
         \caption{Reference image}
         \label{fig:usecaseanalysis_clean}
     \end{subfigure}
     \hfill
    \begin{subfigure}[b]{0.32\linewidth}
         \centering
         \includegraphics[width=0.55\textwidth]{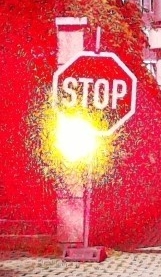}
          \caption{\ac{HDR} inactive}
         \label{fig:usecaseanalysis_blinding_lin}
     \end{subfigure}
     \hfill
    \begin{subfigure}[b]{0.32\linewidth}
         \centering
         \includegraphics[width=0.55\textwidth]{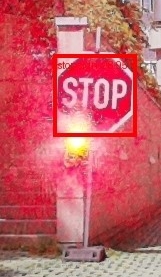}
         \caption{\ac{HDR} active}
         \label{fig:usecaseanalysis_blinding_hdr}
     \end{subfigure}
  \caption{Defensive analysis: blinding attack and \ac{HDR} imaging as a possible mitigation for the region of interest.}
  \label{fig:usecaseanalysis}
  \afterfigspace
\end{figure}
\begin{figure}[t!]
\begin{center}
  \includegraphics[width=0.85\linewidth]{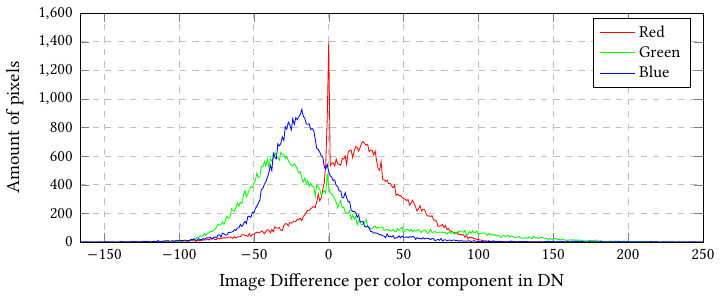}
  \caption{Histogram of the difference of Figures~\ref{fig:usecaseanalysis_blinding_lin} and \ref{fig:usecaseanalysis_blinding_hdr}.}
  \label{fig:hist_usecase}
 \afterfigspace
\end{center}
\end{figure}

For our test, we placed a real stop sign 10 meters away from our testbed and positioned the laser of a commercially available laser rangefinder (Leica Disto A5) directly below the stop sign, aimed straight into the camera. The benign reference image is shown in \autoref{fig:usecaseanalysis_clean}. The correct stop sign goes undetected when the camera is blinded and \ac{HDR} is inactive. \autoref{fig:hist_usecase} shows the histogram of the difference between \autoref{fig:usecaseanalysis_blinding_lin} (without \ac{HDR}) and \autoref{fig:usecaseanalysis_blinding_hdr} (with \ac{HDR}) of the post-\ac{ISP} image as one possibility to compare the images with the low-level analysis tool. While the red color channel shows a high amount of equal pixel values, representing the saturated pixels of the laser beam, there is a local maximum at an image difference of approx. 25 DN. By contrast, the green and blue channels show a maximum difference at a negative value. This means that the non-\ac{HDR} image (\autoref{fig:usecaseanalysis_blinding_lin}) shows a more intense red color in the analyzed region, while green and blue are stronger in the \ac{HDR} image (\autoref{fig:usecaseanalysis_blinding_hdr}). This stronger color differentiation can explain why the stop sign is detected in the \ac{HDR} image. The further the maxima of the individual color components drift apart, the higher the impact of the \ac{HDR} since the color components will be strongly separated.

\subsection{Attack Use Case Analysis}
\label{sec:attackUseCaseAnalysis}

\begin{figure}[t]
\centering
    \begin{subfigure}[b]{0.3\linewidth}
         \centering
         \includegraphics[width=0.7\textwidth]{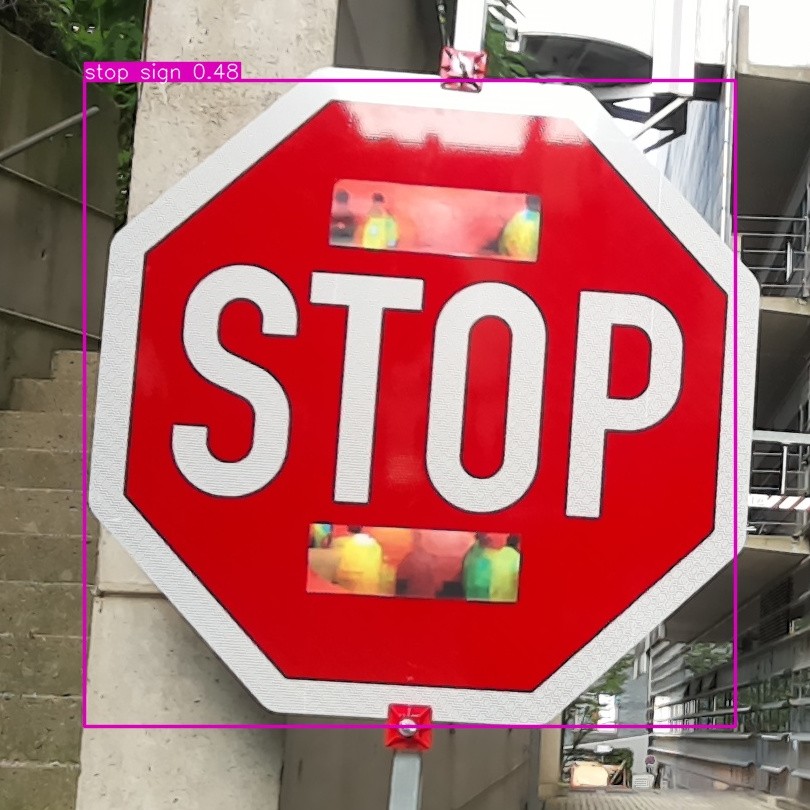}
            \caption{Physical world sticker}
         \label{fig:usecaseanalysis2_orig}
     \end{subfigure}
     \hfill
    \begin{subfigure}[b]{0.3\linewidth}
         \centering
         \includegraphics[width=0.7\textwidth]{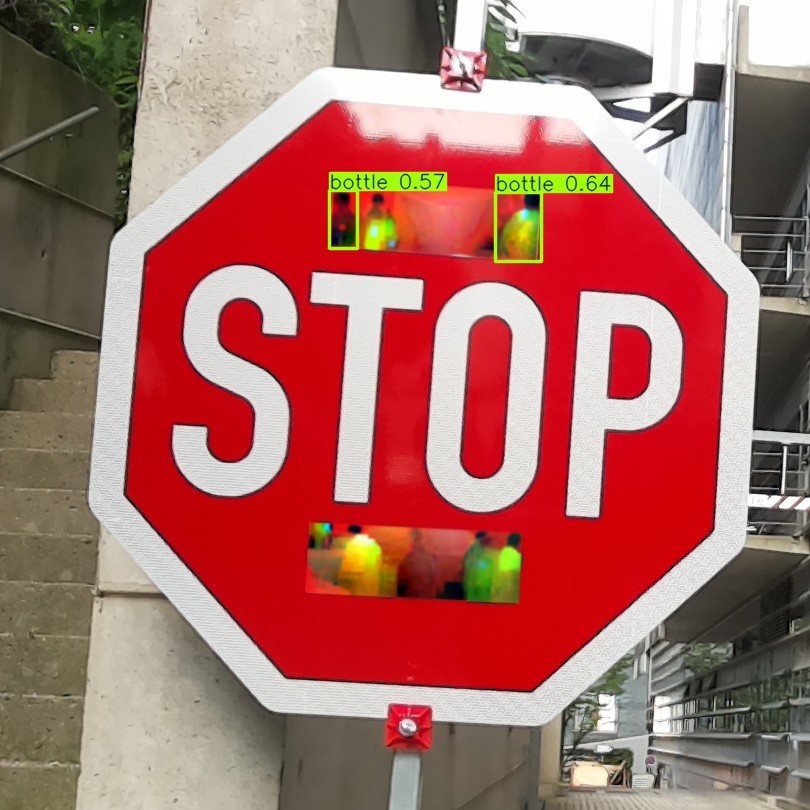}
          \caption{Original digital sticker}
         \label{fig:usecaseanalysis2_digital}
     \end{subfigure}
     \hfill
    \begin{subfigure}[b]{0.3\linewidth}
         \centering
         \includegraphics[width=0.7\textwidth]{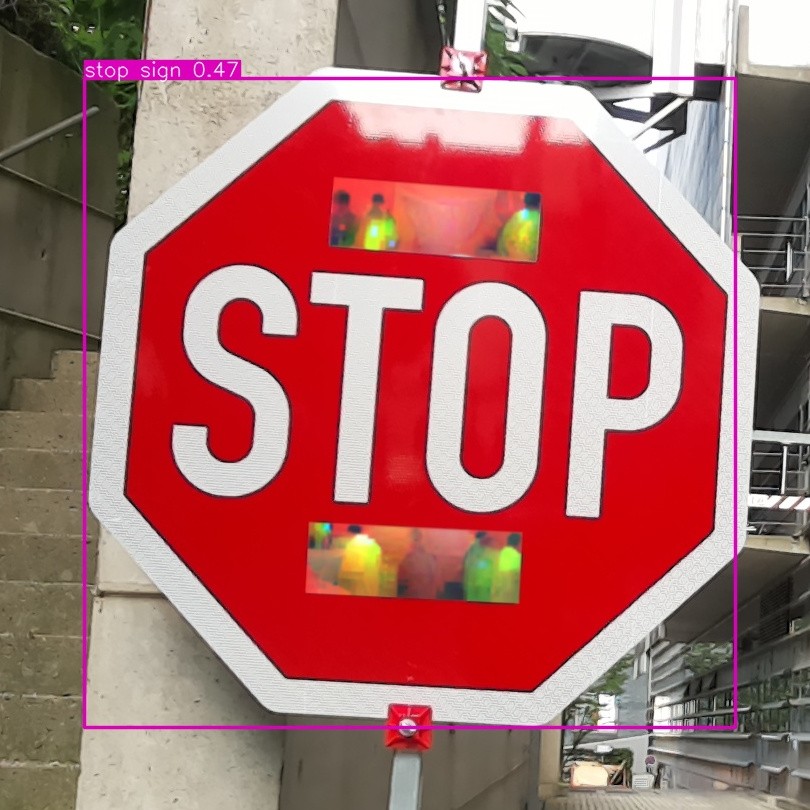}
           \caption{Improved digital sticker}
         \label{fig:usecaseanalysis2_digital_improved}
     \end{subfigure}
  \caption{Attack use case analysis: Physical adversarial sample compared to digital representation and improved digital representation for the region of interest.}
  \label{fig:usecaseanalysis2}
  \afterfigspace
\end{figure}
\begin{figure}[t]
	\begin{center}
		\includegraphics[width=0.7\linewidth]{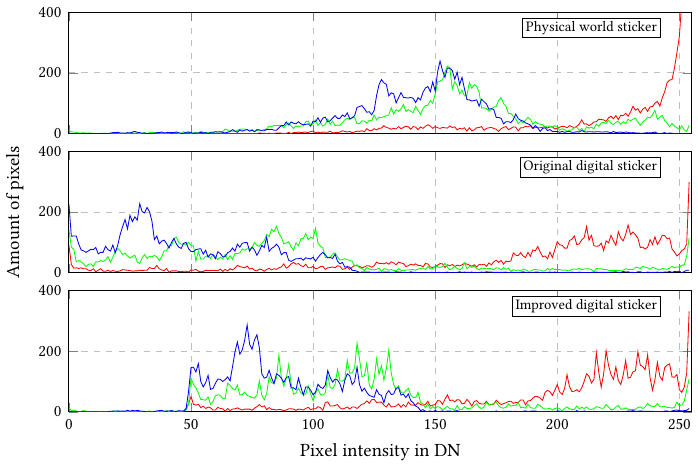}
		\vspace{-1em}
		\caption{Histograms for the upper stickers in \autoref{fig:usecaseanalysis2}.}
		\label{fig:usecase2_low_level}
		\afterfigspace
	\end{center}
\end{figure}

In another use case analysis, we use PICT to demonstrate how low-level pipeline analysis can bridge the gap between digitally designed attacks and their physical-world counterparts. For this test, we use adversarial patches of~\cite{song_physical_2018}, a variant of~\cite{eykholt_robust_2018}, which applies an adversarial patch to a stop sign to evade detection by the YOLO object detector. We select this use case since some attacks~\cite{eykholt_robust_2018, duan_adversarial_2021} provide a digital simulation of the physical stickers to create and optimize the adversarial patch. In the first step, we capture a real-world stop sign with the applied printed sticker as shown in \autoref{fig:usecaseanalysis2_orig}. We use this image to replace the printed sticker with a digital overlay, depicted in \autoref{fig:usecaseanalysis2_digital}. By performing a low-level analysis, we can see the impact of the overall image processing pipeline and iteratively improve the digital sticker to represent the printed one more accurately in \autoref{fig:usecaseanalysis2_digital_improved}.

Although the real-world hiding attack leads to a low confidence of 0.48 in \autoref{fig:usecaseanalysis2_orig}, the original digital sticker results in a more successful hiding attack in \autoref{fig:usecaseanalysis2_digital}. \autoref{fig:usecase2_low_level} shows in the upper plot that the histograms of the physical and digital representation of the sticker deviate significantly. By adjusting the brightness of the digital sticker based on the low-level analysis, we can reach a confidence of 0.47 for the improved digital sticker, which is close to the confidence of the physical sticker. As shown in the lower plot of \autoref{fig:usecase2_low_level}, the intensity follows the physical sticker closer.

\subsection{Processing Pipeline Use Case Analysis}
\label{sec:imageProcessingPipelineUseCaseAnalysis}

As previously shown, many attacks were evaluated using mobile phone cameras~\cite{cao_invisible_2021, avidan_all_2022, kohler_they_2021, sayles_invisible_2021, long_side_2023, liu_cross-task_2024, yan_rolling_2022, li_light_2020, wu_illumination_2021, jan_connecting_2019, zhao_seeing_2019, ji_poltergeist_2021, eykholt_robust_2018}. While convenient, they have image sensors, optics, and \ac{ISP} configurations that differ significantly from automotive cameras, leading to attack evaluations that may not transfer to real systems. We use PICT to compare the effect of adversarial RP\textsubscript{2} patches~\cite{eykholt_robust_2018} when captured with an automotive-like camera and a mobile phone. For our test, we use an iPhone SE 2020~\cite{gsmarena_apple_2020}, with a comparable camera as in the original work~\cite{eykholt_robust_2018, gsmarena_apple_2025, google_nexus_2025}. We craft the stickers to perform a targeted attack against the Inception v3 classifier model~\cite{szegedy_rethinking_2016} to misclassify a "traffic sign" into a "spindle," an arbitrarily chosen target class.
\begin{figure}[t]
\centering
\hspace{0.01\linewidth}
    \begin{subfigure}[b]{0.3\linewidth}
         \centering
         \includegraphics[width=0.7\textwidth]{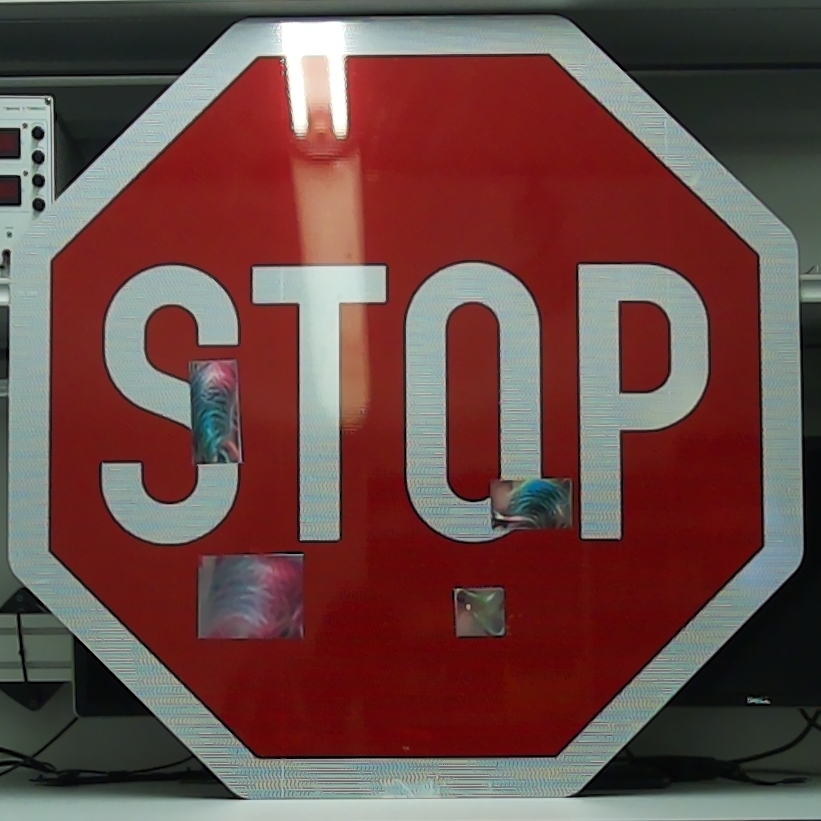}
            \caption{PICT, classified as a "traffic sign."}
         \label{fig:usecaseanalysis3_pict}
     \end{subfigure}
     \hspace{0.15\linewidth}
    \begin{subfigure}[b]{0.3\linewidth}
         \centering
         \includegraphics[width=0.7\textwidth]{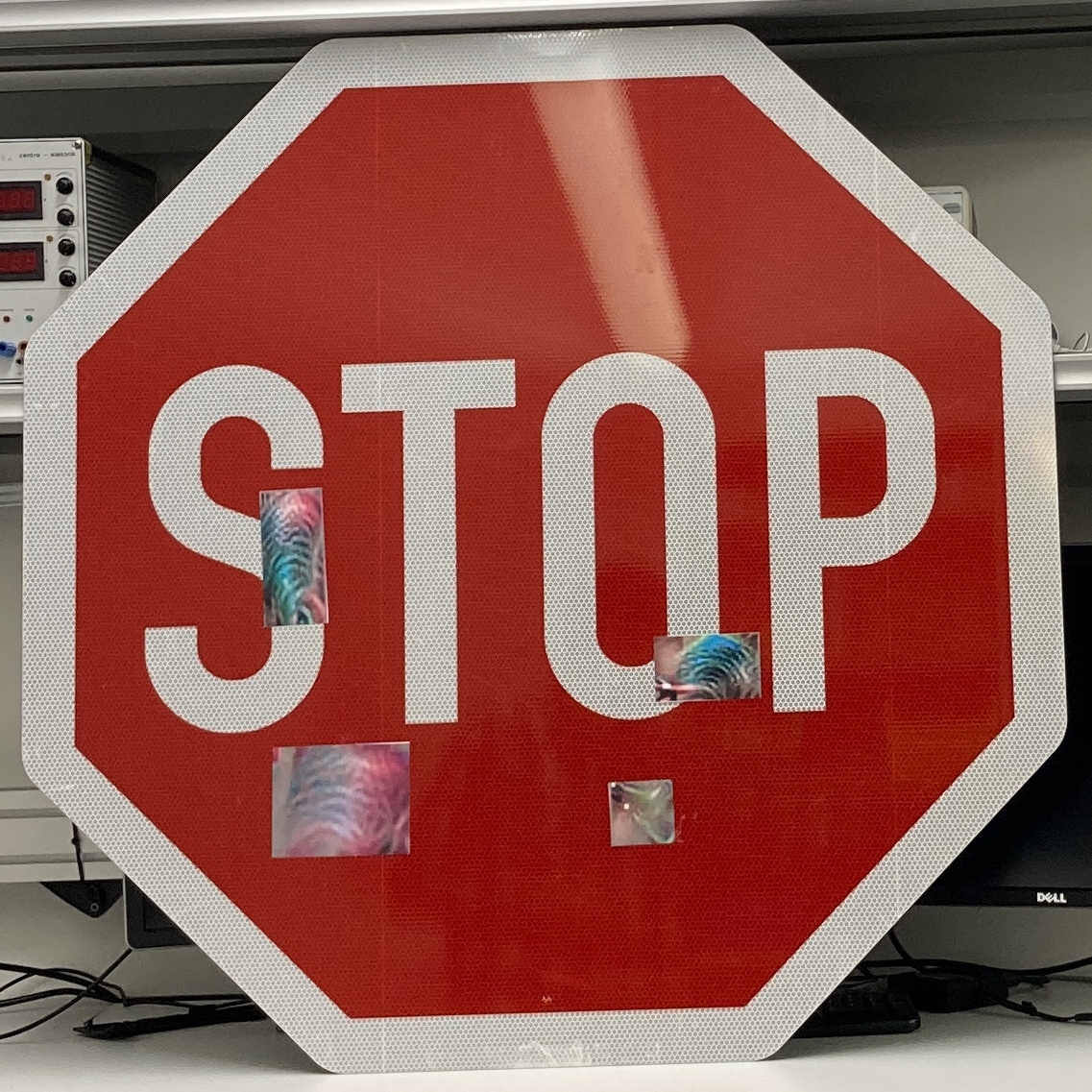}
          \caption{iPhone, classified as a "moving van."}
         \label{fig:usecaseanalysis3_iphone}
     \end{subfigure}
     \hfill
    \caption{Impact analysis: While the patched stop sign from the iPhone is repeatedly misclassified, images captured with PICT are correctly classified by Inception v3.}
    \label{fig:usecaseanalysis3}
    \afterfigspace
\end{figure}
As shown in \autoref{fig:usecaseanalysis3}, we captured ten images of the adversarially patched stop sign with PICT and an iPhone SE 2020. Although both image sensors capture images with $\approx$12MP, iPhone images from \autoref{fig:usecaseanalysis3_iphone} are constantly misclassified with an average confidence of 0.40. By contrast, PICT images from \autoref{fig:usecaseanalysis3_pict} are only misclassified in three out of ten images with an average confidence of 0.18. We repeated the measurements in a darker environment, confirming these results: All captured iPhone images are misclassified, while images captured with PICT are not.
This use case analysis shows that different cameras with different configurations can yield significantly different attack results, despite capturing images under identical environmental conditions.
\section{Conclusion}
\label{sec:conclusion}

In this \ac{SoK}, we investigated security-related research on all layers of the image processing pipeline. We classified the attacks on this pipeline by adapting the automotive security standard ISO 21434. Additionally, we surveyed robustness-related research on the pipeline and identified six main classes of work. By analyzing both security- and robustness-related research, we aimed to take a first step towards closing the existing research gap between these domains and identified the need for a testing environment. With our proposed testbed PICT, we enable researchers to modify several components on all layers of the pipeline and analyze the system impact of individual component changes, allowing further research on security and robustness. Three use case analyses show the benefit of our testbed and highlight that the two research domains must be considered in unison. Future work on the five identified research gaps can help make the overall system more secure and robust.

\begin{acks}
We thank Markus Nickl for creating the 3D model of the camera mount of our testbed. This work is supported by the European Union-funded project CyberSecDome (Agreement No.: 101120779). We gratefully acknowledge the support provided by the U.S. Department of Transportation (DOT) through the National Center for Transportation Cybersecurity and Resiliency (TraCR) grant \#~No.69A3552344812-2027534.
\end{acks}

\bibliographystyle{ACM-Reference-Format}
\bibliography{bibliography_short}

\appendix
\section{Research Methodology}
\label{app:researchmethod}
As a literature basis for this \ac{SoK}, we searched in common search engines, such as "Google Scholar" for scientific literature published in the security and robustness domains of the image processing pipeline, following the snowballing system~\cite{wohlin_guidelines_2014}. For both domains, we consider work irrespective of its publication date but limited our search to research that applies to autonomous vehicles, as this is a given constraint of our \ac{SoK}.

As part of the security-related work, we included the work based on the following criteria:
\begin{itemize}
    \item The work exploits working principles of (parts of) the image processing pipeline of \autoref{fig:teaser}.
    \item The work represents a pure attack, an attack with proposed mitigation, or a pure defense concept.
\end{itemize}

Similarly, for the robustness-related work, we included it based on the following criteria:
\begin{itemize}
    \item The work improves and/or analyzes (parts of) components in the image processing pipeline concerning both image quality and more robust \ac{CV} applications.
    \item The work represents either an analysis of the existing system or provides improvements on the robustness.
\end{itemize}

We explicitly exclude the following research:

\begin{itemize}
    \item Work that shows more generic physical adversarial samples. We refer to existing surveys for interested readers~\cite{guesmi_physical_2023, wang_survey_2023, wang_does_2023, wei_physical_2024, wei_visually_2023}.
    \item Work that works on more abstract levels for both attacks~\cite{akhtar_advances_2021, szegedy_intriguing_2014, su_one_2019} and defenses~\cite{madry_towards_2018, xie_mitigating_2018}.
    \item Work that focuses on improving machine learning models irrespective of autonomous vehicles or irrespective of the image processing pipeline. For interested readers, we refer to existing surveys~\cite{liu_deep_2020, zhang_towards_2020, zhao_object_2019}.
\end{itemize}
\section{Risk Analysis of Camera Attacks}
\label{app:riskAnalysisCameraAttacks}
To evaluate the risk of the analyzed attacks on the image processing pipeline of \autoref{tab:SecurityLayer}, we used the steps given by the "Threat analysis and risk assessment methods" in ISO/SAE 21434:2021(E)~\cite{international_organization_for_standardization_isosae_2021}. Here, we want to provide additional insights into the available categories and the necessary steps to evaluate the risk.

\subsection{Asset Identification}
In the first step, the asset must be identified. Since we analyze existing attacks on the image processing pipeline depicted in \autoref{fig:teaser}, we use the specified layers and the linked work as the asset identification. If further details, such as the exact component, are required, we refer to \autoref{tab:SecurityLayer}, where we specify the affected component. This asset identification only represents the affected component and is not required to calculate the risk.

\subsection{Threat Scenario Identification}
Since we reference existing research on attacks, the respective linked work gives the threat scenario identification. We do not consider the threat scenario identification separately, but we classify the existing attacks into eight classes, as shown in \autoref{tab:SecurityLayer}.

\subsection{Impact Rating}
The impact of the attack must be linked to an impact category, which can be severe \mbox{(\hspace{-0.2em}\impact{4}\hspace{-0.2em})}, major \mbox{(\hspace{-0.2em}\impact{3}\hspace{-0.2em})}, moderate \mbox{(\hspace{-0.2em}\impact{2}\hspace{-0.2em})}, or negligible \mbox{(\hspace{-0.2em}\impact{1}\hspace{-0.2em})}. The standard recommends using four criteria resulting in the impact category: Impact rating on safety damage, financial damage, operational damage, and privacy damage. Each category can have the same criteria as mentioned before. Since the standard explicitly excludes the specification of weightings of each impact category, we considered only the impact rating on safety damage and operational damage since they have the most drastic impact on the safety of the people and the driving experience. \autoref{tab:impactRatingIso} shows the safety and operational impact criteria. Additionally, the standard gives the freedom to add additional impact categories. We added the impact category "Accuracy," specifying the targeted accuracy of the attack, with targeted attacks rated as having a higher impact.

\begin{table}
    \centering
    \caption{Criteria for impact rating based on ISO 21434~\cite{international_organization_for_standardization_isosae_2021}.}
    \label{tab:impactRatingIso}
    \resizebox{0.85\linewidth}{!}{%
    \begin{tabular}{|C{1.4cm}|L{2.8cm}|L{3.1cm}|}\hline
         \textbf{Rating} & \textbf{Safety Impact} & \textbf{Operational Impact} \\ \hline
         Severe \impact{4} & Life-threatening or fatal injuries & Loss or impairment of core vehicle function \\ \hline
         Major \impact{3} & Severe and life-threatening injuries & Loss or impairment of important vehicle function \\ \hline
         Moderate \impact{2} & Light or moderate injuries & Partial degradation of vehicle function \\ \hline
         Negligible \impact{1} & No injuries & No or non-perceivable impairment \\ \hline
    \end{tabular}}
\end{table}

\subsection{Attack Path Analysis}
Similar to asset identification, the existing research work provides attack path analysis. To provide more details on the attack path, we specify the attack entry point in \autoref{tab:comparisonSecurity}, which can be different from the layer where the attack is actually effective. The attack path analysis provides additional information on the attack.

\subsection{Attack Feasibility Rating}
To rate the feasibility of the attack, four ratings are specified in the standard: High \mbox{(\HighFeasibility{})}, Medium \mbox{(\MediumFeasibility{})}, Low \mbox{(\LowFeasibility{})} and Very low \mbox{(\VeryLowFeasibility{})}. Three different methods are proposed to rate the feasibility of which we selected the attack potential-based approach. This approach specifies five core parameters:

\begin{enumerate}
    \item \textit{Elapsed time}: Time to identify a vulnerability, develop and apply a mitigation. Since most attacks exploit hardware characteristics, distorting the actual meaning of this parameter, we excluded it from the calculation.
    \item \textit{Specialist expertise}: Capability necessary by the attack to perform a successful attack. As shown in \autoref{tab:coreParamSpecialistExpertise}, it can be Layman (\Layman{}), Proficient (\prof{}), Expert (\expert{}) or Multiple Experts (\mexpert{}).

    \begin{table}[t]
    \centering
    \caption{Core parameters for attack feasibility rating.}
    \begin{subtable}{.48\linewidth}
        \caption{Specialist Expertise}
        \label{tab:coreParamSpecialistExpertise}
        \resizebox{1\linewidth}{!}{%
        \begin{tabular}{|c|c|c|}\hline
            \textbf{Enumerate} & \textbf{Symbol} &\textbf{Value} \\ \hline
            Layman &\Layman{} & 0 \\
            Proficient &\prof{} & 3\\ 
            Expert & \expert{} & 6\\
            Multiple Experts &\mexpert{} & 8\\
            \hline         
        \end{tabular}}
    \end{subtable}
    \begin{subtable}{.49\linewidth}
        \caption{Knowledge}
        \label{tab:coreParamKnowledge}
        \resizebox{1\linewidth}{!}{%
        \begin{tabular}{|c|c|c|}\hline
        \textbf{Enumerate} & \textbf{Symbol} &\textbf{Value} \\ \hline
            Black-box & \WhiteCircle{} & 0\\
            Gray-box &  \GrayCircle{} & 5\\ 
            White-box & \BlackCircle{} & 11\\
        \hline         
        \end{tabular}}
    \vspace{1em}
    \end{subtable}
    \vspace{1em}
    \begin{subtable}{\linewidth}
        \centering
        \caption{Window of Opportunity}
        \label{tab:coreParamWoO}
        \begin{tabular}{|c|c|c|}\hline
            \textbf{Enumerate} & \textbf{ISO terminology} &\textbf{Value} \\ \hline
            $<100\textrm{m}$ &  Unlimited & 0 \\
            $<10\textrm{m}$ &  Easy & 1 \\
            $<1\textrm{m}$ &  Moderate & 4 \\
            $<0.5\textrm{m}$ &  Moderate & 4 \\
            $<0.1\textrm{m}$ &  Difficult / None & 10 \\
            Remote & Difficult / None & 10 \\
            \hline         
        \end{tabular}
    \end{subtable}
    \begin{subtable}{\linewidth}
        \centering
        \caption{Equipment}
        \label{tab:coreParamEquipment}
        \begin{tabular}{|c|c|c|}\hline
            \textbf{Enumerate} & \textbf{Description} & \textbf{Value}\\ \hline
            Standard & Readily available tools & 0 \\
            Specialized & Available with moderate effort & 4 \\
            Bespoke & Specially produced equipment & 7 \\
            Multiple Bespoke & Various specialized equipment & 9 \\
            \hline         
        \end{tabular}
    \end{subtable}
\end{table}
\begin{table}[t]
\begin{center}
\caption{Risk matrix based on impact and feasibility.}
\label{tab:RiskMatrix}
\resizebox{0.85\linewidth}{!}{%
  \begin{tabular}{|c|C{1.4cm}|C{1.3cm}|C{1.3cm}|C{1.3cm}|C{1.3cm}|C{1.3cm}|} \cline{3-6}
   \multicolumn{2}{c|}{} & \multicolumn{4}{c|}{\textbf{Feasibility}} \\ \cline{3-6}
   \multicolumn{2}{c|}{} & Very Low & Low & Medium & High \\
   \multicolumn{2}{c|}{} & \mbox{\VeryLowFeasibility{}} & \mbox{\LowFeasibility{}} & \mbox{\MediumFeasibility{}} & \mbox{\HighFeasibility{}} \\ \hline
   \multirow{8}{*}{\rotatebox[origin=c]{90}{\textbf{Impact}}}& Severe & \cellcolor{risk1} & \cellcolor{risk3} & \cellcolor{risk4} & \cellcolor{risk5} \\
    & \mbox{\impact{4}} & \multirow{-2}{*}{\cellcolor{risk1}1} & \multirow{-2}{*}{\cellcolor{risk3}3} & \multirow{-2}{*}{\cellcolor{risk4}4} & \multirow{-2}{*}{\cellcolor{risk5}5} \\ \cline{2-6}
    & Major & \cellcolor{risk1} & \cellcolor{risk2} & \cellcolor{risk3} & \cellcolor{risk4} \\
    & \mbox{\impact{3}} & \multirow{-2}{*}{\cellcolor{risk1}1} & \multirow{-2}{*}{\cellcolor{risk2}2} & \multirow{-2}{*}{\cellcolor{risk3}3} & \multirow{-2}{*}{\cellcolor{risk4}4} \\ \cline{2-6}
    & Moderate & \cellcolor{risk1} & \cellcolor{risk1} & \cellcolor{risk2} & \cellcolor{risk3} \\
    & \mbox{\impact{2}} & \multirow{-2}{*}{\cellcolor{risk1}1} & \multirow{-2}{*}{\cellcolor{risk1}1} & \multirow{-2}{*}{\cellcolor{risk2}2} & \multirow{-2}{*}{\cellcolor{risk3}3} \\ \cline{2-6}
    & Negligible & \cellcolor{risk1} & \cellcolor{risk1} & \cellcolor{risk1} & \cellcolor{risk1} \\
    & \mbox{\impact{1}} & \multirow{-2}{*}{\cellcolor{risk1}1} & \multirow{-2}{*}{\cellcolor{risk1}1} & \multirow{-2}{*}{\cellcolor{risk1}1} & \multirow{-2}{*}{\cellcolor{risk1}1} \\ \hline
  \end{tabular}
  }
\end{center}
\end{table}
    \item \textit{Knowledge}: Required knowledge of the item or component. In the given research work, the knowledge can be derived from the assumed threat model, which can be either white-box \mbox{(\WhiteCircle{})}, gray-box \mbox{(\GrayCircle{})} or black-box \mbox{(\BlackCircle{})} as depicted in \autoref{tab:coreParamKnowledge}.
    \item \textit{Window of Opportunity}: Description of the access conditions to successfully perform an attack. In contrast to the standard, we use the distance of the attack that was evaluated in the original research work. We assign the same numerical values to these categories as given in the standard. An overview is given in \autoref{tab:coreParamWoO}.
    \item \textit{Equipment}: Tools (hardware and/or software) required to perform the attack. As shown in \autoref{tab:coreParamEquipment}, it ranges from standard to multiple bespoke.
\end{enumerate}

The overall attack feasibility rating results as a sum of these individual core parameters
\begin{enumerate*}[before=\unskip{: }, itemjoin={{, }}, itemjoin*={{, and }}, label={\textbf{\arabic*)}}]
    \item High ($0 - 13$)
    \item Medium ($14 - 19$)
    \item Low ($20 - 24$)
    \item Very Low ($25 - 38$).
\end{enumerate*}

\subsection{Risk Value Determination}
Based on the impact rating and the attack feasibility rating, the risk value shall be a value between \colorbox{risk1}{\makebox[1em]{\strut\textbf{1}}} and \colorbox{risk5}{\makebox[1em]{\strut\textbf{5}}}, with \colorbox{risk1}{\makebox[1em]{\strut\textbf{1}}} representing the lowest risk. The standard does not define a method for how the individual parts are used to calculate the overall risk. It provides two examples, namely risk matrices or risk formulas. We include the rating of impact and feasibility for each attack and calculate the risk value by using the risk matrix of \autoref{tab:RiskMatrix}.

\end{document}